\documentclass[seceq]{ptptex}

\usepackage{amsfonts,amssymb}

\def\eqref#1{(\ref{#1})}

\def\text#1{\hbox{#1}}
\def\paragraph#1{\noindent{\bf #1}}

\def\linebreak{\hfill\break}

\def\tend{\rightarrow}

\def\therefore{\mbox{\setbox0=\hbox{X}\hbox{$\ldotp$}\raise0.7\ht0\hbox{$\ldotp$}\hbox{$\ldotp$}} \quad }
\def\because{\mbox{\setbox0=\hbox{X}\raise0.7\ht0\hbox{$\ldotp$}\hbox{$\ldotp$}\raise0.7\ht0\hbox{$\ldotp$}}\kern0pt }

\def\bm#1{\boldsymbol{#1}}

\def\Frac(#1/#2){\left(\frac{#1}{#2}\right)}

\def\Eq#1{\begin{equation} #1 \end{equation}}
\def\Eqr#1{\begin{eqnarray} #1 \end{eqnarray}}

\def\Eqrsub#1{\begin{subequations}
\Eqr{#1}\end{subequations}}
\def\Eqrsubl#1#2{\begin{subequations}\label{#1}
\Eqr{#2}\end{subequations}}
\def\Bitm{\begin{itemize}}
\def\Eitm{\end{itemize}}
\newtheorem{theorem}{Theorem}

\def\ZR{{{\mathbb Z}}}

\def\RF{{{\mathbb R}}}

\def\Isom{{\rm Isom}}

\def\OG{\hbox{\it O}}

\def\IO{\hbox{\it IO}}

\def\dS{\hbox{dS}}
\def\AdS{\hbox{AdS}}

\title{Vacuum Branes in $D$-Dimensional Static Spacetimes with 
Spatial Symmetry $\IO(D-2)$, $\OG(D-1)$ or $\OG_+(D-2,1)$ }

\author{Hideo Kodama}

\inst{Yukawa Institute for Theoretical Physics\\
Kyoto University, Kyoto 606-8502, Japan}

\abst{
In this paper, we give a complete classification of vacuum branes, 
i.e., everywhere umbilical time-like hypersurfaces whose extrinsic 
curvature is a constant multiple of the induced metric, 
$K_{\mu\nu}=\sigma g_{\mu\nu}$, in $D$-dimensional static spacetimes 
with spatial symmetry $G(D-2,K)$, where $G(n,K)$ is the isometry 
group of an $n$-dimensional space with constant sectional curvature 
$K$. $D\ge4$ is assumed. It is shown that all possible 
configurations of a brane are invariant under an isometry subgroup 
$G(D-3,K')$ for some $K'\ge K$. In particular, configurations of a 
brane with $\sigma\not=0$ are always $G(D-2,K)$ invariant, except 
for those in five special one-parameter families of spacetimes. 
Further, such 
$G(D-2,K)$-invariant configurations are allowed only in spacetimes 
whose Ricci tensors are isotropic in the two planes orthogonal to each 
$G(D-2,K)$-orbit, or for special values of $\sigma$, which do not 
exist in generic cases. On the basis of these results, we prove the 
non-existence of a vacuum brane with black hole geometry in static 
bulk spacetimes with spatial symmetry $G(D-2,K)$. We also discuss 
mathematical implications of these results.}

\begin{document}

\maketitle

\section{Introduction}

Recently, braneworld models in which our universe is realized as a 
boundary or a subspace, called a brane, of a higher-dimensional 
spacetime have been actively studied as possible new universe models based 
on higher-dimensional unified theories. For example, in the models 
proposed by Randall and Sundrum,
\cite{Randall.L&Sundrum1999,Randall.L&Sundrum1999a} 
our universe is modeled as a boundary of a 5-dimensional vacuum 
spacetime with negative 
cosmological constant whose extrinsic curvature $K_{\mu\nu}$ is 
related to the energy-momentum tensor $T_{\mu\nu}$ of the universe 
by $\kappa_5^2T_{\mu\nu}=2(K^\mu_\nu-K^\lambda_\lambda 
\delta^\mu_\nu)$, where $\kappa_5^2$ is the gravitational coupling 
constant in a 5-dimensional spacetime called a bulk spacetime. In 
this model, when the universe is empty, the extrinsic curvature 
becomes a constant multiple of the induced metric $g_{\mu\nu}$: 
\Eq{
K_{\mu\nu}=\sigma g_{\mu\nu}. 
\label{UmbilicCondition}
}
From the Gauss equation, the brane geometry becomes Minkowskian only 
for some non-zero value of $\sigma$ determined by the cosmological 
constant when the bulk spacetime is an anti-de Sitter spacetime.  
In more realistic models, the bulk spacetime is not a vacuum and 
contains various fields, such as a radion stabilizer and moduli 
fields.%
\cite{Ovrut.B2002A,Goldberger.W&Wise1999,DeWolfe.O&&2000,%
Gibbons.G&Kallosh&Linde2001}
Even in these models, the extrinsic curvature of the brane 
satisfies the condition 
\eqref{UmbilicCondition}, provided that the universe is empty and a 
$\ZR_2$ symmetry is imposed at the brane.

The viability of the braneworld scenario has been investigated with
respect to various aspects, such as the behavior of local gravity and
gravitational waves, the existence and the behavior of FRW universe
models, and the behavior of cosmological perturbations.%
\cite{Randall.L&Sundrum1999a,Garriga.J&Tanaka2000,%
Tanaka.T&Montes2000,Giddings.S&Katz&Randall2000,%
Kudoh.H&Tanaka2001,Vollick.D2001,Kraus.P1999,%
Cline.J&Grojean&Servant1999,Ida.D2000,%
Kodama.H&Ishibashi&Seto2000,Koyama.K&Soda2000,Kodama.H2000A,%
Langlois.D&&2001,Koyama.K&Soda2002} 
However, it is still unknown what kind of black 
holes will be produced by the gravitational collapse of stars. A natural 
starting point for the investigation of this problem is to find 
a regular solution that represents a black hole in the vacuum brane 
and has a localized horizon in the bulk spacetime, but no such 
solution has yet been found, although various suggestive arguments 
have been presented.%
\cite{Chamblin.A&Hawking&Reall2000,Emparan.R&Horowitz&Myers2000,%
Emparan.R&Horowitz&Myers2000a}
From the uniqueness theorems for black holes in 4-dimensional 
spacetimes (for reviews, see Refs. 
\citen{Hawking.S&Ellis1973B,Heusler.M1996B,%
Chrusciel.P1994B,Chrusciel.P1996}) 
and that for static black holes in higher dimensions,%
\cite{Gibbons.G&Ida&Shiromizu2002A} one may naively expect that the 
Schwarzschild-anti-de Sitter solution provides such a solution. 
However, it does not give a black hole solution in the brane, because 
this solution does not contain a non-spherically symmetric vacuum 
brane, as pointed out by Chamblin, Hawking and 
Reall.\cite{Chamblin.A&Hawking&Reall2000} This result is consistent 
with the analysis based on the C-metric solution in the 
4-dimensional bulk spacetime.%
\cite{Emparan.R&Horowitz&Myers2000,Emparan.R&Horowitz&Myers2000a}

One purpose of the present paper is to generalize this result by
Chamblin, Hawking and Reall and to show that there exists no vacuum
brane configuration with black hole geometry in a $D$-dimensional
static spacetime with spatial symmetry $G(D-2,K)$, where $G(n,K)$ is
the isometry group of an $n$-dimensional space with constant sectional
curvature $K$. We assume that $D\ge4$, but do not impose the Einstein
equations on the metric of the bulk spacetime, in order to make the
results applicable to more general braneworld models. We also show
that the existence of a vacuum brane strongly constrains the geometry
of the bulk spacetime.

The second purpose of this paper is rather mathematical. In mathematical 
terminology, a vacuum brane with $\sigma=0$ represents a totally 
geodesic hypersurface. Hence, our analysis completely determines all 
possible totally geodesic time-like hypersurfaces in static 
spacetimes with the assumed spatial symmetry. In contrast, a vacuum 
brane with $\sigma\not=0$ is only geodesic with respect to null 
geodesics. However, such a hypersurface belongs to a more general 
class called `everywhere umbilical hypersurfaces', which are defined 
as hypersurfaces whose extrinsic curvature is proportional to the 
induced metric.%
\footnote{The proportionality coefficient $\sigma$ for an everywhere 
umbilical hypersurface need not be constant on the hypersurface in 
general. However, it is constrained to be constant by the Codazzi 
equation when the embedding bulk space(-time) has a constant 
curvature, as shown in \S5.} 
Concerning such hypersurfaces in Euclidean spaces, there is a 
well-know theorem that an everywhere umbilical 
hypersurface is a hyperplane or a hypersphere.%
\cite{Kobayashi.S&Nomizu1963B} 
We extend this theorem to everywhere umbilical surfaces in non-flat
spacetimes including constant curvature spacetimes. For this purpose,
we do assume no symmetry property on a brane initially, and instead
determine what kind of symmetry the brane should have. We also extend
the well-known rigidity theorem on hypersurfaces in Euclidean spaces
to non-flat spacetimes.

The paper is organized as follows. In the next section, we 
give a general formula expressing the extrinsic curvature of a 
hypersurface in a generic spacetime in terms of a coordinate 
relation specifying the hypersurface, and derive basic equations 
describing a vacuum brane in the class of spacetimes considered in 
the present paper. Then, in \S\S\ref{sec:Sol:S=R} and 
\ref{sec:Sol:S=1}, we solve these basic equations to find all 
possible solutions for the brane configuration as well as 
constraints on the geometry of the bulk spacetime. In 
\S\ref{sec:classification}, we classify isometry classes of the 
solutions found in these two sections and clarify their geometrical 
features. Finally, \S\ref{sec:discussion} is devoted to a 
summary and discussion. Some geometrical formulas and proofs of 
auxiliary theorems used in the text are given in Appendices.

\section{Basic equations}

In this section, we derive the basic equations that determine 
possible configurations of a vacuum brane in a static bulk of 
dimension $D$($\ge4$) with spatial symmetry $\OG(D-1)$, 
$\OG_+(D-2,1)$ or $\IO(D-2)$.

\subsection{Extrinsic curvature of a hypersurface}

We first derive a general expression for the extrinsic curvature of a 
hypersurface $\Sigma$ in a $D$-dimensional general spacetime, 
\Eq{
ds_D^2=\tilde g_{MN}dx^M dx^N.
}
(In this section, capital Latin indices $L,M,\cdots,R$ run over 
$0,\cdots , D-1$.) Let $y^M$ be another coordinate system such that 
$y^{D-1}=0$ on the hypersurface $\Sigma$ and $y^\mu$ 
($\mu=0,1,\cdots,D-2$) gives an intrinsic coordinate system of 
$\Sigma$. Then, the Christoffel symbol $\tilde 
\Gamma'{}^L_{MN}$ in the $y$-coordinates is related to the Christoffel 
symbol $\tilde\Gamma^L{}_{MN}$ in the $x$-coordinates by
\Eq{
\frac{\partial x^L}{\partial y^P}\tilde\Gamma'{}^P_{QR}
=\tilde\Gamma^L_{MN}\frac{\partial x^M}{\partial y^Q}
\frac{\partial x^N}{\partial y^R}
+\frac{\partial^2 x^L}{\partial y^Q \partial y^R}.
}

Now, let $n^M$ be the components of the unit normal to $\Sigma$ in the 
$x$-coordinates and $n'{}^M$ be those in the $y$-coordinates. Then, from 
this relation and the coordinate transformation law of the normal 
vector, the extrinsic curvature of $\Sigma$ in the $y$-coordinates is 
expressed as
\Eqr{
& K_{\mu\nu}
&=-\tilde\nabla_\mu n'_\nu=\tilde\Gamma'{}^M_{\mu\nu}n'_M \nonumber\\
&&=n_L\left(\tilde\Gamma^L_{MN}\partial_\mu x^M \partial_\nu x^N
+\partial_\mu\partial_\nu x^L\right),
\label{K:general}}
where $\partial_\mu$ is the derivative with respect to the 
$y^\mu$-coordinate, and it is understood that the value on $\Sigma$ 
is taken on the right-hand side of this equation.%
\footnote{Note that, in spite of its non-tensorial appearance, the 
right-hand side of \eqref{K:general} is invariant under an arbitrary 
change of the coordinates $x^M$ and transforms as a 2nd-rank tensor 
on $\Sigma$ under a change of the $y^\mu$-coordinates.} 
Further, the components of the normal vector are expressed as
\Eq{
n_L=\pm \tilde N^{-1} \epsilon_{LM_1\cdots M_{D-1}}
\frac{\partial x^{M_1}}{\partial y^0}
\cdots
\frac{\partial x^{M_{D-1}}}{\partial y^{D-2}},
\label{normal:general}}
where $\tilde N$ is a positive normalization constant determined by the 
condition $\tilde g^{MN}n_Mn_N$ $=1$. 

In subsequent applications, we specify the hypersurface by picking out 
one coordinate, say $x^{D-1}$, and expressing $x^{D-1}$ as a 
function of the other coordinates, $X(x^0,\cdots,x^{D-2})$. In this 
case, it is natural to choose the $y$-coordinates such that 
$x^\mu=y^\mu$ for $\mu=0,\cdots,D-2$ and 
$x^{D-1}=y^{D-1}+X(y^0,\cdots,y^{D-2})$. Then, by taking the 
direction of $n^M$ so that $n_{D-1}>0$, the expression for $n_L$ is 
simplified as
\Eq{
n_{D-1}=\tilde N^{-1},\quad
n_\mu=-\tilde N^{-1}\partial_\mu X,
}
and $\tilde N$ is given by
\Eq{
\tilde N^2=\tilde g^{D-1\,D-1}-2\tilde g^{\mu\,D-1}\partial_\mu X 
+\tilde g^{\mu\nu}\partial_\mu X \partial_\nu X.
}
Further, \eqref{K:general} reduces to
\Eqr{
& \tilde N K_{\mu\nu}=
& \partial_\mu\partial_\nu X + \tilde \Gamma^{D-1}_{\mu\nu}
-\partial_\lambda X \tilde\Gamma^\lambda_{\mu\nu}
+2\tilde\Gamma^{D-1}_{D-1(\mu}\partial_{\nu)}X
-2\partial_\lambda 
X\tilde\Gamma^\lambda_{D-1(\mu}\partial_{\nu)}X\nonumber\\
&& +(\tilde\Gamma^{D-1}_{D-1\,D-1}-\partial_\lambda X 
\tilde\Gamma^\lambda_{D-1\,D-1})
\partial_\mu X\partial_\nu X.
}
Similarly, the induced metric $g_{\mu\nu}$ of $\Sigma$ is expressed in 
terms of $\tilde g_{MN}$ and $X$ as
\Eq{
g_{\mu\nu}=\tilde g_{\mu\nu}+2\tilde g_{D-1(\mu}\partial_{\nu)}X
+\tilde g_{D-1\,D-1}\partial_\mu X \partial_\nu X.
}
%

\subsection{Equation for the brane configuration}

Now, we apply the formulas derived above to a vacuum brane in a 
$D$-dimensional static bulk spacetime with spatial symmetry 
$G(D-2,K)$, whose metric is given by
\Eqr{
&&  ds_D^2=-U(R)dT^2+\frac{dR^2}{W(R)}+S(R)^2 d\sigma_{D-2}^2; 
\label{bulkgeometry:symmetric}\\
&&  d\sigma_{D-2}^2=\gamma_{ij}dz^idz^j
=d\chi^2+\rho(\chi)^2d\Omega_{D-3}^2,
\label{metric:M_K}}
where $d\sigma_{D-2}^2$ is the metric of a $(D-2)$-dimensional constant 
curvature space $M^{D-2}_K$ with sectional curvature $K$ ($K=0,\pm1$), 
$d\Omega_{D-3}^2=\hat \gamma_{AB}d\theta^A d\theta^B$ is the metric of 
the $(D-3)$-dimensional unit sphere, and $\rho(\chi)$ is given by 
\Eq{
\rho(\chi)=\left\{\begin{array}{ll}
\sin\chi; & \OG(D-1)\text{-symmetric}\ (K=1),\\
\sinh\chi; & \OG_+(D-2,1)\text{-symmetric}\ (K=-1),\\
\chi; & \IO(D-2)\text{-symmetric}\ (K=0).
\end{array}\right.
\label{rho:def}}
Note that the expression \eqref{bulkgeometry:symmetric} has a gauge
freedom corresponding to an arbitrary reparametrization of the
$R$-coordinate. We fix this gauge freedom by setting $S(R)=R$ when
$S'\not\equiv0$, and $W(R)=1$ and $S(R)=S_0$=const when
$S'\equiv0$. We do not impose the Einstein equations on the bulk
geometry in the present paper. Therefore, at this point, $W(R)$ in the
case $S(R)=R$ and $U(R)$ are arbitrary functions that are positive for
some range of $R$.

In the present paper, we only consider a vacuum brane  of the 
Randall-Sundrum type. Hence, the brane configuration is determined by 
the condition \eqref{UmbilicCondition}. In general, the brane 
configuration can be represented as a level surface of some function 
$F$ on the spacetime as
\Eq{
\Sigma: F(T,R,z^i)=0.
}
Since a brane is a time-like hypersurface, we can assume that 
$\partial_R F\not\equiv0$ or $\partial_i F\not\equiv0$. In the first 
case, the brane configuration can be written $R=R(T,\bm{z})$, and 
from the general formulas in the previous subsection and the 
formulas for the Christoffel symbols in Appendix \ref{appendix:bulkgeometry:formulas},  
\eqref{UmbilicCondition} gives the following equations for 
$R(T,\bm{z})$:
%
\Eqrsubl{eqs:R}{
&& R_{TT}-\left(\frac{W'}{2W}+\frac{U'}{U}\right)R_T^2+\frac{1}{2}WU'
=\sigma N \left(-U+\frac{R_T^2}{W}\right),
\label{eqs:R:TT}\\
&& 
R_{Ti}-\left(\frac{S'}{S}+\frac{U'}{2U}+\frac{W'}{2W}\right)R_TR_i=\sigma 
N \frac{R_TR_i}{W},
\label{eqs:R:Ti}\\
&& 
D_iD_jR-\left(\frac{2S'}{S}+\frac{W'}{2W}\right)R_iR_j
-SS'W\gamma_{ij}=\sigma N \left(\frac{R_iR_j}{W}+S^2\gamma_{ij}\right),
\label{eqs:R:ij}\\
&& N^2=W-\frac{R_T^2}{U}+\frac{R_iR_j\gamma^{ij}}{S^2}.
\label{eqs:R:N}
}
Here, the subscripts $T$ and $i$ on $R$ represent differentiations
with respect to $T$ and $z^i$, respectively, $D_i$ is the covariant
derivative with respect to $\gamma_{ij}$ on $M_K^{D-2}$, and the prime
denotes differentiation with respect to the argument of the
corresponding function.

In the case $\partial_i F\not\equiv0$, the brane 
configuration can be written $\chi=\chi(T,R,\bm{\theta})$ with an 
appropriate choice of the coordinate system of the $(D-2)$-dimensional 
constant curvature space. For this choice,  
\eqref{UmbilicCondition} gives the following equations for 
$\chi(T,R,\bm{\theta})$:
%
\Eqrsubl{eqs:chi}{
&& \chi_{TT}+SS'W\chi_R\chi_T^2-\frac{WU'}{2}\chi_R
=\sigma N\left(-\frac{U}{S}+S\chi_T^2\right),
\label{eqs:chi:TT}\\
&& 
\chi_{TR}+\left(-\frac{U'}{2U}+\frac{S'}{S}\right)\chi_T
+SS'W\chi_R^2\chi_T=\sigma N S \chi_T\chi_R,
\label{eqs:chi:TR}\\
&& 
\chi_{RR}+\left(\frac{W'}{2W}+\frac{2S'}{S}\right)\chi_R+SS'W\chi_R^3
=\sigma N\left(\frac{1}{SW}+S\chi_R^2\right),
\label{eqs:chi:RR}\\
&& 
\chi_{TA}-\frac{\rho'}{\rho}\chi_T\chi_A+SS'W\chi_R\chi_T\chi_A=\sigma
 NS\chi_T\chi_A,
\label{eqs:chi:TA}\\
&& 
\chi_{RA}-\frac{\rho'}{\rho}\chi_R\chi_A+SS'W\chi_R^2\chi_A=\sigma 
NS\chi_R\chi_A,
\label{eqs:chi:RA}\\
&& \hat D_A\hat D_B \chi -2\frac{\rho'}{\rho}\chi_A\chi_B
+SS'W\chi_A\chi_B\chi_R+(SS'W\rho^2\chi_R-\rho\rho')\hat\gamma_{AB}
\nonumber\\
&& \qquad =N\sigma S(\chi_A\chi_B+\rho^2\hat\gamma_{AB}),
\label{eqs:chi:AB}\\
&& N^2=1-\frac{S^2\chi_T^2}{U}+S^2W\chi_R^2
+\frac{1}{\rho^2}\chi_A\chi_B\hat\gamma^{AB}.
\label{eqs:chi:N}
}
Here, the subscripts $T$, $R$ and $A$ for $\chi$ represent the 
derivatives with respect to $T$, $R$ and $\theta^A$, respectively, and 
$\hat D_A$ is the covariant derivative with respect to the metric 
$\hat\gamma_{AB}$ on the unit sphere $S^{D-3}$.

\section{Solutions in spacetimes with 
$S=R$}\label{sec:Sol:S=R}

In this section, we solve the equations for the brane configuration 
derived in the previous section for the bulk metric with $S(R)=R$. 
We divide the problem into the case in which the brane configuration 
has non-trivial $R$-dependence and the case in which the brane 
configuration is independent of $R$. The former case is described by 
\eqref{eqs:R}, while the latter case is determined by 
\eqref{eqs:chi} with $\chi=\chi(T,\bm{\theta})$.  

\subsection{$R=R(T,\bm{z})$-type configurations}

In order to solve \eqref{eqs:R}, we replace the variable $R$ by the 
new variable $r=r(R)$ defined by
\Eq{
dr=\frac{dR}{R\sqrt{W}}
}
and introduce $V(R)$ and $Z(R)$ defined by
\Eq{
V=\frac{\sqrt{U}}{R},\quad Z=\frac{\sqrt{W}}{R}.
}
In terms of these variables, \eqref{eqs:R:Ti} is written
\Eq{
r_{Ti}=\left(\frac{V_r}{V}+RZ+\frac{\sigma N}{Z}\right)
r_T r_i.
}
As shown in Appendix \ref{appendix:DTDiu}, this equation implies 
that $r$ depends on $\bm{z}=(z^i)$ through some function $X(\bm{z})$ as 
$r=r(T,X(\bm{z}))$. Hence, from this point, we regard  $r$ as a 
function of the two variables $T$ and $X$. Then, \eqref{eqs:R} can be  
written
\Eqrsubl{eqs:r}{
&& r_{TT} +\frac{V_r}{V}(V^2-2r_T^2)
+\left(RZ+\frac{\sigma N}{Z}\right) 
(V^2- r_T^2)=0,
\label{eqs:r:TT}\\
&& r_{TX}
=\left(\frac{V_r}{V}+RZ+\frac{\sigma N}{Z}\right)
r_Tr_X,
\label{eqs:r:TX}\\
&& r_X D_iD_j X+ \left[r_{XX}
-r_X^2\left(RZ+\frac{\sigma N}{Z}\right)\right] X_iX_j
=\left(RZ+\frac{\sigma N}{Z}\right)\gamma_{ij},
\label{eqs:r:ij}\\
&& \frac{N^2}{R^2Z^2}=1-\frac{r_T^2}{V^2}+r_X^2(DX)^2,
\label{eqs:r:N}
}
where $(DX)^2=X_iX_j\gamma^{ij}$.

\subsubsection{Configurations represented as $R=R(T)$}

First, we treat the special case in which $r_i\equiv0$, i.e., 
$r_X\equiv0$. In this case, \eqref{eqs:r:ij} gives $\sigma N=-RZ^2$, 
which is equivalent to $\sigma<0$ and 
\Eq{
\frac{1}{Z^2}\left(\frac{r_T^2}{V^2}-1\right)=-\frac{1}{\sigma^2}.
\label{r_X=0:r_T}}
Differentiation of this equation with respect to $T$ gives
\Eq{
r_T\left[r_{TT}+\frac{V_r}{V}(V^2-2r_T^2)+(r_T^2-V^2)\left(\frac{V_r}
{V}-\frac{Z_r}{Z}\right)\right]=0.
}
Furthermore, in the present case,  \eqref{eqs:r:TT} reads
\Eq{
r_{TT}+\frac{V_r}{V}(V^2-2r_T^2)=0.
\label{r_X=0:r_TT}}
Comparing these two equations, we find that $V_r/V=Z_r/Z$ for 
$r_T\not=0$. Hence, $U$ is proportional to $W$ in the range of $R$ 
in which $R_T\not=0$. Since $U$ can be multiplied by an arbitrary 
positive constant by rescaling the time variable $T$, this implies 
that we can set $U=W$. Note that this condition is equivalent to the 
condition that $R^a_b$ is proportional to $\delta^a_b$ for 
$a,b=R,T$. Under this condition, the brane configuration equations 
are equivalent to the condition $\sigma<0$ and the single equation 
\eqref{r_X=0:r_T}, which is expressed in terms of the original 
variable as 
\Eq{
R_T^2=U^2\left(1-\frac{U}{\sigma^2R^2}\right).
\label{Sol:R_i=0:R_T/=0}}
This type of brane configuration represents a FRW-type cosmological 
model. On the other hand, if $r_T\equiv0$ (i.e., the configuration 
is represented by $R=$const), \eqref{r_X=0:r_T} and 
\eqref{r_X=0:r_TT} give the two conditions
\Eq{
\frac{U'}{U}=\frac{2}{R},\quad
\sigma^2=\frac{W}{R^2}.
\label{Sol:R_i=0:R_T=0}}
Although no general constraint on $U$ or $W$ arises in this case, 
values of $R$ satisfying these equations, if they exist, form a 
discrete set when $U$ is not proportional to $R^2$ in any finite 
range of $R$, and the value of $\sigma$ also becomes discrete.

\subsubsection{Configurations with $R_i\not\equiv0$}

Next, we consider the case in which $r_i\not\equiv0$, i.e., 
$r_X\not\equiv0$ and $X_i\not\equiv0$. In this case, 
\eqref{eqs:r:ij} is written
\Eq{
D_iD_j X =\alpha \gamma_{ij} + \beta D_iX D_jX,
\label{eq:X}}
where
\Eq{
\alpha =\frac{1}{r_X}\left(RZ+\frac{\sigma N}{Z}\right),\quad 
\beta=-\frac{r_{XX}}{r_X}+r_X^2\alpha.
}
Since $r$ is a function of $T$ and $X$, it follows from 
\eqref{eqs:r:TX} that $\alpha$ can also be written 
$\alpha=\alpha(T,X(\bm{z}))$. Hence, $\beta$ also depends on
$\bm{z}=(z^i)$ only through $X(\bm{z})$. Further, since $X$ is
independent of $T$, and since $X_iX_j$ is linearly independent of
$\gamma_{ij}$ under the condition $X_i\not\equiv0$, $\alpha$ and
$\beta$ should be $T$-independent. Therefore, \eqref{eq:X} is an
equation of the type discussed in Appendix \ref{appendix:DDu}. As
shown there, by replacing $X$ by some monotonic function of $X$, we
can always make $\beta$ vanish. Since we have the freedom to make such
a redefinition of $X$ in the present case, we can assume that $X$
satisfies the condition $\beta=0$. Then, from the argument in Appendix
\ref{appendix:DDu}, $\alpha$ must be of the form $-KX+\alpha_0$, where
$\alpha_0$ is a constant. Thus,
\eqref{eqs:r:ij} is equivalent to the set of equations
\Eqrsubl{eqs:X}{
&& D_iD_j X=(\alpha_0-KX)\gamma_{ij},
\label{eqs:X:ij}\\
&& RZ+\frac{\sigma N}{Z}=(\alpha_0-KX)r_X,
\label{eqs:X:Z}\\
&& r_{XX}=r_X^3(\alpha_0-KX).
\label{eqs:X:r}
}
%


Now, we show that the consistency of the equations for the brane 
configuration requires $W$ to have the form 
\Eq{
W=K-\lambda R^2,
}
where $\lambda$ is a constant. First, from \eqref{eqs:X:Z} and 
\eqref{eqs:X:r}, \eqref{eqs:r:TX} can be written 
\Eq{
r_{TX}=\left(\frac{V_rr_X}{V}+\frac{r_{XX}}{r_X}\right)r_T.
}
This implies that $r_T/Vr_X$ is independent of $X$, and hence 
$r_T$ can be written 
\Eq{
r_T=\dot g(T) Vr_X,
\label{eq:r_T}}
in terms of some function $g(T)$. Next, as shown in Appendix 
\ref{appendix:DDu}, $(DX)^2$ is written 
\Eq{
(DX)^2=c-KX^2+2\alpha_0 X,
\label{eq:(DX)^2}}
where $c$ is a constant. Further, the integration of 
\eqref{eqs:X:r} with respect to $X$ yields
\Eq{
-\frac{1}{r_X^2}=2\alpha_0 X -KX^2-f(T)^2-\dot g(T)^2+c,
\label{eq:r_X}}
where $f(T)^2$ is an arbitrary function of $T$. Inserting these 
expressions into \eqref{eqs:r:N}, we obtain
\Eq{
\frac{N^2}{R^2Z^2}=f^2 r_X^2,
}
which implies that $f$ can be taken to be positive definite. Hence, 
\eqref{eqs:X:Z} can be written 
\Eq{
\frac{\alpha_0-KX}{R}-\frac{Z}{r_X}=\sigma 
\frac{r_X}{|r_X|}f=\pm\sigma f.
\label{eq:Z}}
Since the right-hand of this equation is independent of $X$,  
differentiation of this equation with respect to $X$ yields
\Eq{
Z_r=-\frac{K}{R}.
}
This equation with $(1/R)_r=-Z$ derived from the definition of $r$ 
leads to $(Z^2-K/R^2)_r=0$, which implies that $(W-K)/R^2$ is 
constant.


Up to this point, we have not used \eqref{eqs:r:TT}. Next, we show 
that the consistency of this equation with the others provides 
additional strong constraints. With the help of \eqref{eqs:X:Z} and 
\eqref{eqs:X:r}, \eqref{eqs:r:TT} can be rewritten 
\Eq{
r_{TT}-\left(\frac{2V_r}{V}+\frac{r_{XX}}{r_X^2}\right)r_T^2 
+V^2\left(\frac{V_r}{V}+\frac{r_{XX}}{r_X^2}\right)=0.
}
Since the sum of the first two terms on the left-hand side of this 
equation is equal to $Vr_X(r_T/Vr_X)_T$, using \eqref{eq:r_T}, it can
be further rewritten as
\Eq{
\ddot g + \frac{V_r}{r_X}+(\alpha_0-KX)V=0.
\label{eq:ddotg}}
Also, differentiating \eqref{eq:r_X} with respect to $T$ 
and eliminating $r_{TX}$ using \eqref{eq:r_T}, we obtain
\Eq{
\dot g\left[\ddot g+\frac{V_r}{r_X}+(\alpha_0-KX)V\right]
=-f\dot f.
}
Thus, the consistency of these two equations requires that $f$ be 
constant. Then, differentiation of \eqref{eq:Z} with respect to $T$ 
gives
\Eq{
0=\pm \sigma \dot f=(-VZ_r+V_rZ)\dot g.
}
From this, it follows that if $\dot g\not\equiv0$, then $V_r/V=Z_r/Z$, 
which implies that $U$ is proportional to $W$. Thus, we find that 
the equations for the brane configuration have a solution 
$R=R(T,\bm{z})$ such that $R_T$ and $R_i$ do not vanish identically  
only when $U=W=K-\lambda R^2$ up to a constant rescaling of $T$, 
i.e., only when the bulk spacetime is either a de Sitter spacetime 
dS$^D$ ($\lambda>0$), an anti-de Sitter spacetime AdS$^D$ 
($\lambda<0$), or a Minkowski spacetime $E^{D-1,1}$.

To summarize the argument up to this point, the original equations 
for the brane configuration \eqref{eqs:R} can be replaced by 
\eqref{eq:r_T}, 
\eqref{eq:r_X}, \eqref{eq:Z}, \eqref{eq:ddotg} and $W=K-\lambda 
R^2$. The integrability condition for the first two equations with 
respect to $r$ is given by the condition that $f$ is a positive 
constant, which will be denoted by $a$ from this point. Further, the 
third is consistent with the others only if $\dot g\equiv 0$ or 
$U=W=K-\lambda R^2$ up to a constant rescaling of $T$. Finally, from 
the argument in Appendix \ref{appendix:DDu}, $X=X(\bm{z})$ is 
determined by \eqref{eqs:X:ij} uniquely up to an isometry of the 
constant curvature space $M_K^{D-2}$ and multiplication by a 
constant that is uniquely determined by $c$ in \eqref{eq:(DX)^2}, 
except in the case $K=-1$ and $c=0$.

In order to proceed further, we must consider the cases 
$R_T\not\equiv0$ and $R_T\equiv0$ separately.

\medskip
\noindent{(A) Solutions with $R_T\not\equiv0$ and 
$R_i\not\equiv0$}

In this case, we can assume that $U=W=K-\lambda R^2$ without loss of 
generality. We consider the three cases $K=0,\pm1$ separately.

\medskip

\noindent{(A-1) The case $K=0$}

In this case, $\lambda<0$ and $V=Z=\sqrt{-\lambda}$. Hence, $r$ is related 
to $R$ as
\Eq{
\sqrt{-\lambda}r=-\frac{1}{R},
}
and the equations for the brane configuration are written
\Eqrsubl{eqs:K=0}{
&& r_T=\sqrt{-\lambda}\dot g r_X,
\label{eqs:K=0:r_T}\\
&& \frac{1}{r_X^2}=-2\alpha_0 X +a^2+\dot g^2 - c,
\label{eqs:K=0:r_X}\\
&& -\frac{1}{r_X}=\alpha_0r\pm \frac{a\sigma}{\sqrt{-\lambda}},
\label{eqs:K=0:Z}\\
&& \ddot g+ \alpha_0 \sqrt{-\lambda}=0.
\label{eqs:K=0:ddotg}
}

As shown in Appendix \ref{appendix:DDu}, if we choose the Cartesian 
coordinate system of $M_0^{D-2}=E^{D-2}$ as $z^i$, $X$ is a linear 
function of $z^i$ for $\alpha_0=0$, while it can be chosen to be 
$(\bm{z}-\bm{z}_0)^2$ for $\alpha_0\not=0$, where $\bm{z_0}$ is a 
constant vector. Hence, these two cases are qualitatively different.

\noindent{i) The case $\alpha_0=0$}:  
First, in the case $\alpha_0=0$, we can normalize $X$ so that 
$c=(DX)^2=1$. Since $\pm$ on the right-hand side of 
\eqref{eqs:K=0:Z} coincides with the sign of $r_X$, it follows that 
$\sigma<0$. Equations \eqref{eqs:K=0:r_X} and \eqref{eqs:K=0:Z} can be
easily integrated, yielding
\Eqr{
&& r=\mp \frac{\sqrt{-\lambda}}{\sigma a}X+h(T),\\
&& \dot g^2=1-\left(1+\frac{\sigma^2}{\lambda}\right)a^2.
}
Inserting the first equation into \eqref{eqs:K=0:r_T}, we obtain 
\Eq{
h=\pm \frac{\lambda}{\sigma a}g,
}
where we have absorbed an integration constant into $g$. Finally, 
\eqref{eqs:K=0:ddotg} determines $g$ to be a linear function of $T$. 
Thus, the general solution for the case $K=\alpha_0=0$ is written
\Eqr{
&& \frac{1}{R}=\frac{\lambda}{\sigma}(\bm{p}\cdot \bm{z} + q 
T)+A;\nonumber\\
&& \lambda \bm{p}^2+q^2=\sigma^2+\lambda,\ \sigma<0,
\label{Sol:R_i/=0:R_T/=0:K=0:alpha=0}
}
where $\bm{z}=(z^i)$ is the Cartesian coordinate system for $E^{D-2}$, 
and $\bm{p}=(p_i)$ is a constant vector.

\noindent{ii) The case $\alpha_0\not=0$}: 
In this case, we can set 
$X=(\bm{z}-\bm{z}_0)^2$ without loss of generality, i.e., $c=0$ and 
$\alpha_0=2$. Then, \eqref{eqs:K=0:ddotg} becomes $\ddot 
g+2\sqrt{-\lambda}=0$, whose general solution is  
\Eq{
g=-\sqrt{-\lambda}(T-T_0)^2+b,
}
where $T_0$ and $b$ are constants. From this expression, 
\eqref{eqs:K=0:r_X} and \eqref{eqs:K=0:Z}, we obtain
\Eq{
\left(\frac{1}{R}\mp \frac{a\sigma}{2}\right)^2
=\lambda (\bm{z}-\bm{z}_0)^2+\lambda^2(T-T_0)^2
-\frac{1}{4}\lambda a^2.
\label{Sol:R_i/=0:R_T/=0:K=0:alpha/=0}}
It can be easily checked that this satisfies \eqref{eqs:K=0:r_T} and 
\eqref{eqs:K=0:r_X}. Hence, this is the general solution for the 
case $K=0$ and $\alpha_0\not=0$.

\medskip

\noindent{(A-2) The case $K=\pm1$}

In this case, we can set $\alpha_0=0$ by a constant shift of $X$. 
Then, the equations for the brane configuration are written
\Eqrsubl{eqs:K/=0}{
&& r_T=\dot g(T) Vr_X,
\label{eqs:K/=0:r_T}\\
&& \frac{1}{r_X^2}=KX^2+a^2+\dot g^2-c,
\label{eqs:K/=0:r_X}\\
&& \frac{KX}{R}+\frac{V}{r_X}=\mp\sigma a,
\label{eqs:K/=0:Z}\\
&& \ddot g+\frac{V_r}{r_X}-KXV=0.
\label{eqs:K/=0:ddotg}
}
From the first equation, it follows that
\Eq{
(X+gV)_T=\left(\frac{1}{r_X}+gV_r\right)r_T.
}
Also, we have the identity
\Eq{
(X+gV)_X=\left(\frac{1}{r_X}+gV_r\right)r_X.
}
The consistency of these two equations requires that $X+gV$ depend 
only on $r$ (or equivalently on $R$). Therefore, we write
\Eq{
X+gV=h(r),\quad
\frac{1}{r_X}+gV_r=h_r,
}
where the second equation automatically follows from the first. 
Hence, \eqref{eqs:K/=0:r_T} is equivalent to the first equation. 
Further, with the help of these equations and $V_r=-K/R$, 
\eqref{eqs:K/=0:Z} is deformed to
\Eq{
\frac{K}{R}h+Vh_r=\mp a\sigma,
\label{eq:h}
}
which provides an equation to determine $h(r)$.

\noindent{i) The case $\lambda\not=0$}: 
First, for the case $\lambda\not=0$, since $(1/R)_r=-Z=-V$ and 
$V_r=Z_r=-K/R$, $h=\mp\frac{a\sigma}{\lambda R}$ gives a special 
solution to \eqref{eq:h} and the general solution is expressed as 
the sum of this special solution and a constant multiple of $V$. The 
latter can be absorbed into the definition of $g$, which has the 
freedom to add an arbibrary constant. Hence, $X$ is expressed as
\Eq{
X=-gV \mp \frac{a\sigma}{\lambda R},
\label{Sol:R_i/=0:R_T/=0:K/=0:lambda/=0}}
which in turn gives
\Eq{
\frac{1}{r_X}=g\frac{K}{R}\pm \frac{a\sigma}{\lambda}V.
}
Inserting this expression into \eqref{eqs:K/=0:r_X} gives 
\Eq{
\dot g^2-K\lambda g^2=c-\left(1+\frac{\sigma^2}{\lambda}\right)a^2.
\label{Sol:R_i/=0:R_T/=0:K/=0:lambda/=0:constraint}}
Further, \eqref{eqs:K/=0:ddotg} reads
\Eq{
\ddot g-K\lambda g=0.
}
This equation is obtained from 
\eqref{Sol:R_i/=0:R_T/=0:K/=0:lambda/=0:constraint} by 
differentiating it with respect to $T$, if $\dot g\not=0$. Hence, 
the solution to the equations for the brane configuration in this 
case is given by \eqref{Sol:R_i/=0:R_T/=0:K/=0:lambda/=0} and 
\eqref{Sol:R_i/=0:R_T/=0:K/=0:lambda/=0:constraint}. The solution to 
the latter equation can be explicitly expressed in terms of a linear 
combination of $\sin(\sqrt{|\lambda|}T)$ and 
$\cos(\sqrt{|\lambda|}T)$ for $K\lambda<0$ or 
$\sinh(\sqrt{|\lambda|}T)$ and $\cosh(\sqrt{|\lambda|}T)$ for 
$K\lambda>0$, if necessary. Further, as described in Appendix 
\ref{appendix:DDu}, $X$ is uniquely determined, up to an isomorphism of 
$M_K^{D-2}$ and multiplication by a constant, as a solution to 
\Eq{
D_iD_jX=-KX\gamma_{ij};\quad
(DX)^2=-KX^2+c.
\label{eq:X:K/=0}
}
%

\noindent{ii) The case $\lambda=0$}: 
In this case, $K$ must be unity, which implies that 
$U=W=1$, and $r$ is related to $R$ by
\Eq{
R=e^r.
}
Further, we can set $X=\cos \chi$ by a constant multiplication and an
isometry, for which $c=1$. Equation \eqref{eq:h} can be easily
integrated to give
\Eq{
X\equiv \cos\chi=-\frac{g}{R}\mp \frac{a\sigma}{2}R.
}
In this case, \eqref{eqs:K/=0:r_X} gives the condition
\Eq{
\mp 2a\sigma g=\dot g^2+a^2-1,
}
and \eqref{eqs:K/=0:ddotg} is written
\Eq{
\ddot g \pm a\sigma=0.
}
Hence, the general solution to the brane configuration equations is 
given by%
\Eqrsubl{Sol:R_i/=0:R_T/=0:K/=0:lambda=0}{
&\sigma\not=0:& 
R\cos\chi=\pm\frac{a\sigma}{2}\left[(T-T_0)^2-R^2\right]+b;\ a^2\mp 
2ab\sigma=1,\\
& \sigma=0: &
R\cos\chi=pT+A;\ p^2<1.
\label{Sol:R_i/=0:R_T/=0:K/=0:lambda=0:sigma=0}
}

\medskip
\noindent{(B) Solutions represented as $R=R(\bm{z})$}

When $R_T\equiv0$, $R$ (and $r$) depends only on $X=X(\bm{z})$, and 
$W$ is still given by $W=K-\lambda R^2$, but $U$ need not coincide 
with $W$. Hence, the brane configuration equations are written
\Eqrsubl{eqs:R_T=0}{
&& \frac{1}{r_X^2}=-2\alpha_0 X+ KX^2+a^2-c,
\label{eqs:R_T=0:r_X}\\
&& \frac{\alpha_0-KX}{R}-\frac{Z}{r_X}=\pm a\sigma,
\label{eqs:R_T=0:Z}\\
&& \frac{V_r}{r_X}+(\alpha_0-KX)V=0.
\label{eqs:R_T:ddot g}
}
Since $r_{XX}/r_X^3=\alpha_0-KX$, the last equation can be easily 
integrated to give
\Eq{
V=\frac{k}{r_X},
\label{eqs:R_T=0:V}}
where $k$ is an integration constant.

\medskip

\noindent{(B-1) The case $K=0$}

As in the corresponding case with $R_T\not\equiv0$, we have 
$Z=\sqrt{-\lambda}$ ($\lambda<0$) and $\sqrt{-\lambda}rR=-1$, while 
solutions for $\alpha_0=0$ and solutions for $\alpha_0\not=0$ are 
qualitatively different.

\noindent{i) The case $\alpha_0=0$}: 
In this case, \eqref{eqs:R_T=0:Z} is written
$r_X=\mp\sqrt{-\lambda}/a\sigma$, which requires $\sigma<0$ and can be
easily integrated to yield
\Eq{
\frac{1}{R}=\mp\frac{\lambda}{a\sigma}X+A,
}
where $A$ is a constant. Inserting this into \eqref{eqs:R_T=0:r_X} 
gives
\Eq{
\lambda c=(\sigma^2+\lambda)a^2.
}
Further, from \eqref{eqs:R_T=0:V}, it follows that $V$ is constant, 
which implies that we can set $U=W$ by a constant rescaling of $T$. 
We can see that this solution is the special case of 
\eqref{Sol:R_i/=0:R_T/=0:K=0:alpha=0} with $q=0$, by choosing the 
representation $X=a\bm{p}\cdot\bm{z}$, for which $c=a^2\bm{p}^2$.

\noindent{ii) The case $\alpha_0\not=0$}: 
In this case, as in the corresponding case 
with $R_T\not=0$, we can set $c=0$ and $\alpha_0=2$, for which 
$X=(\bm{z}-\bm{z}_0)^2$ up to an isometry of $E^{D-2}$. 
Equation \eqref{eqs:R_T=0:Z} can be integrated to yields
\Eq{
r^2+X=\mp\frac{a\sigma}{\sqrt{-\lambda}}r+b,
}
where $b$ is an integration constant. This constant is determined by 
\eqref{eqs:R_T=0:Z} as
\Eq{
b=\frac{\sigma^2+\lambda}{4\lambda}a^2.
}
Hence, we obtain
\Eq{
X\equiv(\bm{z}-\bm{z}_0)^2=\frac{1}{\lambda}\left(\frac{1}{R}\mp 
\frac{a\sigma}{2}\right)^2+\frac{a^2}{4}.\ (\sigma\not=0)
\label{Sol:R_i/=0:R_T=0:K=0:alpha/=0}}
From this expression and \eqref{eqs:R_T=0:V}, $U=R^2V^2$ is 
determined up to a constant factor as
\Eq{
U=\left(1\mp \frac{a\sigma}{2}R\right)^2.
\label{Sol:R_i/=0:R_T=0:K=0:alpha/=0:U}
}
Hence, $U$ is not proportional to $W$.

\medskip

\noindent{(B-2) The case $K\not=0$}

As in the corresponding case with $R_T\not\equiv0$, we can set 
$\alpha_0=0$. Then, from \eqref{eqs:R_T=0:r_X}, rewritten as 
\Eq{
\left(\frac{KX}{R}+\frac{Z}{r_X}\right)^2
-K\left(ZX + \frac{1}{Rr_X}\right)^2
=-\lambda (a^2-c),
}
and \eqref{eqs:R_T=0:Z}, we obtain
\Eq{
ZX + \frac{1}{Rr_X}=b;\ b^2=K[(\sigma^2+\lambda)a^2-\lambda c].
\label{eqs:R_T=0:r_X:1}}
%

\noindent{i) The case $\lambda\not=0$}: 
In this case,  \eqref{eqs:R_T=0:r_X:1} and 
\eqref{eqs:R_T=0:Z} can be regarded as linear equations for $X$ and 
$1/r_X$. Solving these equations and using \eqref{eqs:R_T=0:V}, we 
obtain
\Eqrsubl{Sol:R_i/=0:R_T=0:K/=0:lambda/=0}{
&& U=(bK\pm a\sigma \sqrt{W})^2,
\label{Sol:R_i/=0:R_T=0:K/=0:lambda/=0:U}\\
&& \lambda X=-b\frac{\sqrt{W}}{R}\mp \frac{a\sigma}{R},\\
&& b^2=K[(\sigma^2+\lambda)a^2-\lambda c],
}
after an appropriate rescaling of $U$. $U$ is proportional to $W$ 
only when $b=0$, for which the solution is the special case of 
\eqref{Sol:R_i/=0:R_T/=0:K/=0:lambda/=0} with $g=0$.

\noindent{ii) The case $\lambda=0$}: 
In this case, $K=1$, $W=1$, and 
$R=e^r$. The integration of \eqref{eqs:R_T=0:Z} yields
\Eq{
X=\mp \frac{a\sigma}{2}R+\frac{b}{R}.
\label{Sol:R_i/=0:R_T=0:K/=0:lambda=0}
}
Also, \eqref{eqs:R_T=0:r_X} is written
\Eq{
c=a^2\mp 2ab\sigma,
\label{Sol:R_i/=0:R_T=0:K/=0:lambda=0:constraint}
}
and \eqref{eqs:R_T=0:V} determines $U$ as
\Eq{
U=\left(b\pm \frac{a\sigma }{2}R^2\right)^2,
\label{Sol:R_i/=0:R_T=0:K/=0:lambda=0:U}
}
after a constant rescaling. Since $c>0$ in this case, $U$ is 
proportional to $W=1$ only when $\sigma=0$, for which the solution 
is the special case of 
\eqref{Sol:R_i/=0:R_T/=0:K/=0:lambda=0:sigma=0} with $p=0$.

To summarize, a static configuration with $R_i\not=0$ is allowed 
only in bulk geometries for which $W=K-\lambda R^2$ and $U=W$ or 
$U$ is given by 
\eqref{Sol:R_i/=0:R_T=0:K=0:alpha/=0:U}($K=0$), 
\eqref{Sol:R_i/=0:R_T=0:K/=0:lambda/=0:U}($b\lambda\not=0$) or  
\eqref{Sol:R_i/=0:R_T=0:K/=0:lambda=0:U}($\lambda=0,b\not=0$).

\subsection{$\chi=\chi(T,\bm{\theta})$-type configurations}

In this case, the equations for the brane configuration are given by 
\eqref{eqs:chi} with $\chi_R\equiv0$. In particular, from 
\eqref{eqs:chi:RR} it follows that only the brane with $\sigma=0$ 
can have this type of configuration. Hence, the brane configuration 
equations reduce to the following set of equations:
\Eqrsubl{eqs:chi:chi_R=0}{
&& \chi_{TT}=0,
\label{eqs:chi:chi_R=0:TT}\\
&& V'\chi_T=0,
\label{eqs:chi:chi_R=0:TR}\\
&& \chi_{TA}-\frac{\rho'}{\rho}\chi_T\chi_A=0,
\label{eqs:chi:chi_R=0:TA}\\
&& \hat D_A\hat D_B \left(\frac{\rho'}{\rho}\right)
=-\frac{\rho'}{\rho}\gamma_{AB},
\label{eqs:chi:chi_R=0:AB}\\
&& N^2=1-\frac{\chi_T^2}{V^2}+\frac{1}{\rho^2}(\hat D\chi)^2.
\label{eqs:chi:chi_R=0:N}
}
In deriving \eqref{eqs:chi:chi_R=0:AB} from \eqref{eqs:chi:AB}, 
we have used the identity
\Eq{
\left(\frac{\rho'}{\rho}\right)'=-\frac{1}{\rho^2},
}
which holds for any  value of $K$.

\subsubsection{Solutions with $\chi_T\equiv0$}

For the static case, the brane configuration is determined by 
\eqref{eqs:chi:chi_R=0:AB}. As shown in Appendix \ref{appendix:DDu}, 
by an appropriate $\OG(D-2)$ transformation, any solution to this 
equation can be written 
\Eq{
\frac{\rho'}{\rho}=a\cos\theta,
\label{Sol:chi_T=0}}
where $a$ is a constant and $\theta$ is the geodesic distance from the
north pole of the $(D-3)$-dimensional unit sphere. Note that for
$K=-1$, we have $a^2>1$, since $\cosh\chi/\sinh\chi>1$. No constraint
on $U$ or $W$ arises.

\subsubsection{Solutions with $\chi_T\not\equiv0$}

In the non-static case, \eqref{eqs:chi:chi_R=0:TR} requires $V'$ to 
vanish. Hence, after an appropriate rescaling of $T$, $U$ is written 

\Eq{
U=R^2.
\label{Sol:chi_T/=0:U}}
Further, from the argument in Appendix \ref{appendix:DTDiu}, 
\eqref{eqs:chi:chi_R=0:TA} implies that $\chi$ depends on $\theta^A$ 
only through some function $M(\theta^A)$, and hence $\rho'/\rho$ is 
written $\rho'/\rho=F(T,M(\theta^A))$. Inserting this expression 
into \eqref{eqs:chi:chi_R=0:AB} and using an argument similar to 
that concerning the $T$-independence of $\alpha$ and $\beta$ in 
\eqref{eq:X}, we find that $F$ can be written $F=A(T)M(\theta^A)$
through an appropriate redefinition of $M(\theta^A)$. Hence, we obtain
\Eq{
\frac{\rho'}{\rho}=A(T)\cos\theta,
}
where $\theta$ is the same as in \eqref{Sol:chi_T=0}. Inserting this 
expression into \eqref{eqs:chi:chi_R=0:TA} and using the identity 
$K\rho^2+(\rho')^2=1$, we find that $K=0$. This implies that the 
bulk metric can be expressed as
\Eq{
ds_D^2=\frac{dR^2}{W(R)}+R^2(-dT^2+d\bm{z}^2).
}
Further, 
\eqref{eqs:chi:chi_R=0:TT} determines $A$ to be $A=1/(aT+b)$, where 
$a$ and $b$ are constant. Hence, the general solution is expressed as
\Eq{
aT+b=\chi \cos\theta;\ a^2<1,
\label{Sol:chi_T/=0}
}
where the condition on $a$ comes from $N^2=(1-a^2)/\cos^2\theta>0$. 
Since $\chi\cos\theta$ can be expressed as $\bm{n}\cdot\bm{z}$ in 
terms of a unit vector $\bm{n}$ in $E^{D-2}$, this solution 
corresponds to a time-like hyperplane in $E^{D-2,1}$ with the 
coordinates $(T,\bm{z})$ on each $R=$const section.

\section{Solutions in spacetimes with $S=S_0$}
\label{sec:Sol:S=1}

In this section, we solve the brane configuration equations for the 
Nariai-type bulk geometry whose metric is given by 
\eqref{bulkgeometry:symmetric} with $W=1$ and $S=S_0$. The arguments
run almost parallel to those in the previous section.

\subsection{$R=R(T,\bm{z})$-type configurations}

In this case, \eqref{eqs:R} reads
\Eqrsubl{eqs:R:Nariai}{
&& R_{TT} -\frac{U'}{U}R_T^2+\frac{1}{2}U'=\sigma 
N(-U+R_T^2),\label{eqs:R:Nariai:TT}\\
&& R_{Ti}-\frac{U'}{2U}R_TR_i=\sigma N R_TR_i,
\label{eqs:R:Nariai:Ti}\\
&& D_iD_j R=\sigma N (R_iR_j+S_0^2\gamma_{ij}),
\label{eqs:R:Nariai:ij}\\
&& N^2=1-\frac{R_T^2}{U}+\frac{R_iR_j\gamma^{ij}}{S_0^2}.
\label{eqs:R:Nariai:N}
}
%

\subsubsection{Configurations represented as $R=R(T)$}

We first look for special solutions with $R_i\equiv0$. From 
\eqref{eqs:R:Nariai:ij}, we see that such a solution exists only for 
$\sigma=0$, and that \eqref{eqs:R:Nariai:TT} is the only non-trivial 
equation. For configurations with $R_T\equiv0$, this equation 
reduces to the equation $U'=0$ for constant values of $R$. On the 
other hand, for configurations with $R_i\not\equiv0$, 
\eqref{eqs:R:Nariai:TT} is equivalent to
\Eq{
\left(\frac{R_T^2}{U^2}-\frac{1}{U}\right)_T=0.
}
Hence, $U(R)$ can be an arbitrary function, and the brane 
configuration is described by a solution to the first-order ordinary 
equation
\Eq{
R_T^2=U(1-AU);\ A>0,
\label{Sol:Nariai:R_i=0}
}
where the condition $A>0$ comes from $N^2=AU>0$. Thus, the general 
solution contains one free parameter in addition to the one 
representing time translation.

\subsubsection{Configurations with $R_i\not\equiv0$}

Next, in the case $R_i\not\equiv0$, \eqref{eqs:R:Nariai:Ti} implies 
that $R$ is written $R=R(T,X(\bm{z}))$ and gives
\Eq{
R_{TX}=\left(\frac{U'}{2U}+\sigma N\right)R_TR_X.
\label{eqs:Nariai:TX}}
Hence, repeating the argument of the previous section, we find 
that \eqref{eqs:R:Nariai:ij} is equivalent to the set of equations
\Eqrsub{
&& D_iD_j X=(\alpha_0-KX)\gamma_{ij},\\
&& (DX)^2=-KX^2+2\alpha_0 X + c,\\
&& S_0^2\frac{\sigma N}{R_X}=\alpha_0-KX,
\label{eqs:Nariai:R_i/=0:constraint}\\
&& \frac{R_{XX}}{R_X^3}=\frac{\alpha_0-KX}{S_0^2}.
\label{eqs:Nariai:R_i/=0:R_XX}
}

By using these equations, \eqref{eqs:Nariai:TX} can be integrated to 
yield
\Eq{
R_T=V R_X \dot g(T),
\label{eqs:Nariai:R_i/=0:R_T}
}
where 
\Eq{
V=\sqrt{U}/S_0,
}
and $g(T)$ is an arbitrary function of $T$. Also, integrating
\eqref{eqs:Nariai:R_i/=0:R_XX} once gives 
\Eq{
-\frac{S_0^2}{R_X^2}=-KX^2+2\alpha_0X+c-f^2-\dot g^2,
\label{eqs:Nariai:R_i/=0:R_X}}
where $f$ is a function of $T$. It can be assumed to be positive 
definite, because $N$ is expressed as 
\Eq{
S_0\frac{N}{R_X}=\pm f.
}
Inserting this expression into \eqref{eqs:Nariai:R_i/=0:constraint}, 
we obtain
\Eq{
\alpha_0 -KX=\pm S_0\sigma f(T),
}
from which it follows that $K=0$. Therefore, from this point, we set 
$S_0=1$ by redefining the Euclidean metric $S_0^2\gamma_{ij}$ as 
$\gamma_{ij}$. Then, the bulk metric is given by
\Eq{
ds_D^2=-U(R)dT^2+dR^2+d\bm{z}^2,
\label{Sol:Naria:R_i/=0:metric}
}
and the spatial section becomes the Euclidean space $E^{D-1}$.

With the help of these equations, \eqref{eqs:R:Nariai:TT} can be 
deformed into
\Eq{
\ddot g+ \frac{1}{R_X^3}(VR_X)_X=0.
\label{eqs:Nariai:R_i/=0:ddotg}}
Also, differentiation of \eqref{eqs:Nariai:R_i/=0:R_X} 
with respect to $T$ yields
\Eq{
-f\dot f=\dot g\left(\ddot g+ \frac{1}{R_X^3}(VR_X)_X\right).
}
Hence, $f$ must be a positive constant $a$, which is related to 
$\alpha_0$ as 
\Eq{
\alpha_0=\pm a\sigma.
\label{eqs:Nariai:R_i/=0:alpha}}
To summarize, the equations for the brane configuration are reduced 
to 
\eqref{eqs:Nariai:R_i/=0:R_T}, \eqref{eqs:Nariai:R_i/=0:R_X} with 
$f=a$ and $K=0$, \eqref{eqs:Nariai:R_i/=0:ddotg}, and 
\eqref{eqs:Nariai:R_i/=0:alpha}.

As in the previous section, by considering $d(X+gV)$, we find that 
$X+gV$ depends only on $R$:
\Eq{
X+gV=h(R).
\label{eq:Nariai:h:def}}
In order to determine $h(R)$ and $g(T)$, we regard $X$ as a function 
of $T$ and $R$. Then, since $X_R=1/R_X$, 
\eqref{eqs:Nariai:R_i/=0:R_X} is written
\Eq{
\dot g^2-(V')^2g^2+2(\alpha_0V+h'V')g+a^2-c-2\alpha_0 h-(h')^2=0.
\label{eqs:Nariai:R_i/=0:R_X:1}
}

\medskip
\noindent{(A) Solutions with $R_T\not\equiv0$ and 
$R_i\not\equiv0$}

Let us first consider the case $\dot g\not\equiv0$. In this case, 
since $T$ and $R$ are independent variables, 
\eqref{eqs:Nariai:R_i/=0:R_X:1} leads to 
\Eqr{
&& V'=\text{const},\ \alpha_0V+h'V'=\text{const},
\label{eq:Nariai:V}\\
&& -a^2+c+2\alpha_0 h+(h')^2=b,
\label{eq:Nariai:h}}
where $b$ is a constant. Solutions to these equations differ 
qualitatively depending on whether $V'\equiv0$ or $V'\not\equiv0$. 

\medskip
\noindent{(A-1) The case $U=1$}

In the case $V'\equiv0$, we can set $U=1$ by rescaling $T$. Hence, 
the bulk metric \eqref{Sol:Naria:R_i/=0:metric} becomes the standard 
Minkowski metric. Since 
\eqref{eq:Nariai:h} implies $h''+\alpha_0=0$, its general solution 
is easily found to be%
\Eq{
h=-\frac{1}{2}\alpha_0 R^2+h_1R+h_0;\quad
b=c-a^2+2\alpha_0 h_0+h_1^2,
\label{eq:Nariai:h:sol}}
where $h_0$ and $h_1$ are integration constants.  Further, 
\eqref{eqs:Nariai:R_i/=0:R_X:1} reads
\Eq{
\dot g^2+2\alpha_0 g=b.
\label{eq:Nariai:R_T/=0:g}}
Since $\dot g\not\equiv0$, \eqref{eqs:Nariai:R_i/=0:ddotg} follows 
from this equation. 

For $\sigma=0$, we have $\alpha_0=0$ from \eqref{eqs:Nariai:R_i/=0:alpha}, 
and $X$ is written $X=\bm{p}\cdot\bm{z}$ with $\bm{p}^2=c$ for a 
Cartesian coordinate system $\bm{z}$. Further, the general solution 
to  \eqref{eq:Nariai:R_T/=0:g} is $\pm\sqrt{b}T+\text{const}$. 
Hence, from \eqref{eq:Nariai:h:def} and \eqref{eq:Nariai:h:sol}, the 
general solution to the brane configuration equations is given by
\Eq{
sT=\bm{p}\cdot\bm{z}+qR+A;\ 
s^2<\bm{p}^2+q^2,
\label{Sol:Nariai:U=1:sigma=0}}
where $s,p_i,q$ and $A$ are constants. This solution represents a 
time-like hyperplane in the Minkowski spacetime $E^{D-1,1}$.

On the other hand, for $\sigma\not=0$, we can set 
$X=(\bm{z}-\bm{z}_0)^2$, for which $\alpha_0=2$ and $c=0$. The 
general solution to 
\eqref{eq:Nariai:R_T/=0:g} now reads 
$g=-(T-T_0)^2-a^2/4+h_0+h_1^2/4$. Hence, the solution to the brane 
configuration equations is given by
\Eq{
(\bm{z}-\bm{z}_0)^2+(R-R_0)^2=(T-T_0)^2+\frac{1}{\sigma^2},
\label{Sol:Nariai:U=1:sigma/=0}}
which represents a time-like hyperboloid in the Minkowski spacetime 
$E^{D-1,1}$.

\medskip
\noindent{(A-2) The case $U=R^2$}

Next, we consider the case $V'\not\equiv0$. In this case, we can set 
$U=R^2$ by rescaling $T$. Hence, the bulk metric 
\eqref{Sol:Naria:R_i/=0:metric} coincides with the Minkowskian metric 
in the Rindler coordinates:
\Eq{
ds_D^2=-R^2dT^2+dR^2+d\bm{z}\cdot d\bm{z}
=-(d(R\sinh T))^2+(d(R\cosh T))^2+d\bm{z}\cdot d\bm{z}.
\label{RindlerMetric}
}

First, we consider a brane with $\sigma=0$. In this case, 
$\alpha_0=0$ from \eqref{eqs:Nariai:R_i/=0:alpha}, and $h=h_1R+h_0$ 
from \eqref{eq:Nariai:V}. We can set $h_1=0$ by redefining $g-h_1$ 
as $g$, from 
\eqref{eq:Nariai:h:def}. Hence, $X$ is written
\Eq{
X=h_0-gR,
}
where $g$ is a solution to
\Eq{
\dot g^2-g^2+a^2-c=0,
}
from \eqref{eq:Nariai:R_T/=0:g}. Thus, the general solution to the 
brane configuration equations is given by
\Eq{
\bm{p}\cdot\bm{z}=qR\cosh T + s R\sinh T+ A;\ 
\bm{p}^2+q^2>s^2.
\label{Sol:Nariai:U=R^2:sigma=0}}
From \eqref{RindlerMetric}, we see that this solution represents a 
time-like hyperplane in the Minkowski spacetime.

Next, for $\sigma\not=0$, by setting $\alpha_0=2$ and $c=0$, we find 
that $h$ is expressed as $h=-R^2+h_0$ after a constant shift of $g$. 
The value of $g$ is determined from
\Eq{
\dot g^2-g^2+a^2-4h_0=0.
}
Hence, the general solution to the equations for the brane 
configuration is expressed as
\Eq{
\bm{z}\cdot\bm{z}+(R\cosh T-A)^2=(R\sinh T-B)^2+\frac{1}{\sigma^2}.
\label{Sol:Nariai:U=R^2:sigma/=0}}
This is a time-like hyperboloid in the Minkowski spacetime, 
represented in the Rindler coordinates.

\medskip
\noindent{(B) Solutions represented as $R=R(\bm{z})$}

Finally, we consider the static case, $\dot g\equiv0$. In this case, 
we can set $g\equiv0$ without loss of generality. Then, $X=h(R)$ and 
$U$ is proportional to $X_R^2=(h')^2$ from 
\eqref{eqs:Nariai:R_i/=0:ddotg}. The solution for $h$ is now given 
by \eqref{eq:Nariai:h:sol} with $b=0$ from 
\eqref{eqs:Nariai:R_i/=0:R_X:1}. It follows from these that 
solutions to the brane configuration equations in the present case 
are simply given by setting $s=0$ in \eqref{Sol:Nariai:U=1:sigma=0} 
for $\sigma=0$ and $A=B=0$ in \eqref{Sol:Nariai:U=R^2:sigma/=0} for 
$\sigma\not=0$.

To summarize, a bulk spacetime with $W=1$ and $S=S_0$ can contain a 
vacuum brane with a configuration $R_i\not\equiv0$ only when it is a 
Minkowski spacetime, and allowed configurations of the brane are 
hyperplanes for $\sigma=0$ and hyperboloids for $\sigma\not=0$.

\subsection{$\chi=\chi(T,\bm{\theta})$-type configurations}

For $S=S_0$ and $\chi_R\equiv0$, it is easy to see that 
\eqref{eqs:chi:RR} leads to $\sigma=0$ and the rest of Eqs. 
\eqref{eqs:chi} coincide with the corresponding equations 
\eqref{eqs:chi:chi_R=0} for $S=R$ if we replace $V$ by $U$. Hence, 
solutions to the brane configuration equations of the type 
$\chi=\chi(T,\bm{\theta})$ are given by two families defined by 
\begin{subequations}
\Eq{
U\ \text{is\ arbitrary},\ K=0,\pm1,\ 
\frac{\rho'}{\rho}=a\cos\theta,
}
and
\Eq{
U=1,\ K=0,\ 
aT+b=\chi\cos\theta; \ a^2<1,
}
\label{Sol:Nariai:chi}
\end{subequations}
\noindent
up to transformations by isometries of the bulk spacetime.  Note 
thatthe former family represents a totally geodesic hypersurface in 
theconstant curvature space $M_K^{D-2}$, as in the case with $S=R$, 
whilethe latter family is a special case of
\eqref{Sol:Nariai:U=1:sigma=0} with $q=0$.

\section{Classification of the bulk geometry and the brane 
configuration}\label{sec:classification}

In this section, we classify the solutions obtained in the previous
two sections by identifying solutions connected by
isometries. Throughout this section, the relation $M^D=(B,F)$
signifies that the bulk geometry of $M^D$ is a warped product of a
base space $B$ and a fibre $F$, i.e.,
\Eq{
ds_D^2= g_{ab}(y)dy^ady^b + f(y)^2 \gamma_{ij}(z)dz^i dz^j.
}

We first group the solutions obtained in 
\S\S\ref{sec:Sol:S=R} and \ref{sec:Sol:S=1} into the 
following four types, mainly according to the symmetry 
property of the bulk geometry. 

\paragraph{Type I}: This type is defined as the set of solutions for 
which the brane configuration is  expressed in one of the special 
forms (A), (B) and (C) described below, in some coordinate system in 
which the bulk metric is written in the form 
\eqref{bulkgeometry:symmetric}.  Except for the $\IO(1)\times G(D-2,K)$ 
symmetry, no additional symmetry is assumed on the bulk geometry, and
hence the bulk spacetime has the structure $M^D=(N^2,M_K^{D-2})$,
where $N^2$ is the static 2-dimensional spacetime.
\begin{itemize}
\item[(A)] Static configurations of a brane with $\sigma=0$ 
expressed as
\Eq{
\frac{\rho'}{\rho}=a\cos\theta.
}
\item[(B)] Configurations with $\sigma\not=0$ of the 
form $R=R(T)$ satisfying
\Eq{
R_T^2=U^2\left(1-\frac{U}{\sigma^2 R^2}\right)\not\equiv0
}
in special bulk geometries with $S=R$ and $U=W$, and 
those of a brane with $\sigma=0$ of the form $R(T)$ satisfying
\Eq{
R_T^2=WU(1-AU)\not=0;\ A>0
}
in bulk geometries with $S=$const. 
\item[(C)] Static brane configurations expressed as $R=$const in 
terms of a solution to 
\Eq{
\frac{U'}{U}=\frac{2S'}{S},\quad
W\left(\frac{S'}{S}\right)^2=\sigma^2.
\label{TypeI-C:condition}}
\end{itemize}

\paragraph{Type II}: This type is defined by the 
condition that the bulk geometry is of the type 
$M^D=(M^{D-1}_\lambda,E^{0,1})$ and the metric is written
\Eq{
ds_D^2=d\sigma_{\lambda,D-1}^2 - U(R)dT^2,
\label{TypeII:metric}}
where $d\sigma_{\lambda,D-1}^2$ is the metric of a $(D-1)$-dimensional
constant curvature space $M_\lambda^{D-1}$ with sectional curvature
$\lambda$. $R$ is a coordinate in this space such that each $R=$const
surface has a constant curvature with the same sign as that of $K$,
and the metric is written in the form
\eqref{bulkgeometry:symmetric}. When $U(R)$ is proportional to
$W(R)=K-\lambda R^2$ for $S=R$ or $U=1$ for $\lambda=0$, the spacetime
becomes a constant curvature spacetime with sectional curvature
$\lambda$. This case is excluded, in order to make this type mutually
exclusive with the type IV. We also exclude bulk geometries for which
$d\sigma_{\lambda,D-1}^2=dR^2+d\bm{z}^2$, in order to make this type
mutually exclusive with the type I-B.

\paragraph{Type III}: This type is defined by the condition 
that the bulk geometry has the symmetry $IO(D-2,1)$ and 
$M^D=(E^1,E^{D-2,1})$. The metric is written
\Eq{
ds_D^2=dv^2+f(v)^2 \eta_{\mu\nu}dx^\mu dx^\nu.
\label{TypeIII:metric}}
When $f(v)$ is proportional to $e^{av}$, the geometry becomes
$E^{D-1,1}$ for $a=0$ and $\AdS^D$ for $a\not=0$. These cases are
excluded in order to make this type mutually exclusive with the type
IV.

\paragraph{Type IV}: This type is defined by the condition 
that the bulk spacetime is a constant curvature spacetime, 
i.e., $M^D=E^{D-1,1}$, dS$^D$ or AdS$^D$. In the 
representation \eqref{bulkgeometry:symmetric}, this type 
corresponds to the cases in which $U=W=K-\lambda R^2$ and 
$S=R$, where $\lambda$ coincides with the sectional 
curvature of the bulk from \eqref{Riemann:bulk}, or $K=0$, 
$U=1$ or $R^2$, $W=1$ and $S=$const. 

The solutions for $R_T\not\equiv0$ and $R_i\not\equiv0$ 
with $S=R$ in \S\ref{sec:Sol:S=R}, and those for 
$R_i\not\equiv0$ or $\chi_T\not\equiv0$ with $S=1$ in 
\S\ref{sec:Sol:S=1} belong to the type IV. The solutions 
with $\chi_T\not\equiv0$ in \S\ref{sec:Sol:S=R} belong to 
the type III. As shown there, these solutions correspond to brane 
configurations with $\sigma=0$ that have the structure 
$\Sigma=(E^1,E^{D-3,1})$, where $E^{D-3,1}$ is an 
arbitrary time-like hyperplane of $E^{D-2,1}$. 
The solutions with $R_T\equiv0$ and 
$R_i\not\equiv0$ in \S\ref{sec:Sol:S=R} belong to the type 
II. The solutions with $\chi_T\equiv0$ and $\chi_R\equiv0$ in 
\S\S\ref{sec:Sol:S=R} and \ref{sec:Sol:S=1} belong to one 
of the types I-A, II, III or IV, depending on the bulk 
geometry. The solutions with 
$R_i=0$ in \S\S\ref{sec:Sol:S=R} and \ref{sec:Sol:S=1} 
belong to either  the type I-B or the type I-C, 
depending on whether $R_T\not\equiv0$ or $R_T\equiv0$.

This correspondence can be rephrased in the following way from the 
point of view of the symmetry of a brane. Let $\IO(1)\times G_0$ be 
a subgroup of $\Isom(M^D)$ such that $G_0$ is isomorphic to 
$G(D-2,K)$. Then, first, if a brane is $\IO(1)\times G_0$ invariant, 
solutions are of the type I-C, and $\sigma$ is restricted to some 
special values for analytic metrics. Second, if a brane is $G_0$ 
invariant but not $\IO(1)$ invariant, solutions are of the type I-B, 
and the bulk geometry is constrained. Third, if a brane is $\IO(1)$ 
invariant but not $G_0$ invariant, solutions are of the type I-A, 
the type III with $\sigma=0$, the type II or the type IV. In the 
latter three cases, the bulk geometry is quite special. Finally, if 
a brane is neither $G_0$ invariant nor $\IO(1)$ invariant, solutions 
are of the type III with $\sigma=0$ or of the type IV.

Here, note that the type I and the other types may not be mutually
exclusive, although the types II, III and IV as well as the types I-B,
II and III are mutually exclusive, as will be shown later. Actually,
we show below that the types III, IV and most geometries of the type
II are subclasses of the type I. Further, solutions belonging to
different subtypes of the type I may be geometrically identical in the
case in which the bulk spacetime allows two different space-time
decomposition of the form
\eqref{bulkgeometry:symmetric}. 

The main purpose of the present section is to make these points clear. 
For this purpose, we use a general geometrical argument concerning
everywhere umbilical hypersurfaces of a constant curvature
space(-time), instead of starting from explicit expressions for the
solutions in special coordinate systems obtained in the previous two
sections.

\subsection{Everywhere umbilical hypersurfaces of a constant 
curvature space(-time)}\label{sec:EUHinCCS}

Let $\tilde M_\lambda$ be a constant curvature space(-time) with 
sectional curvature $\lambda$ and $\Sigma$ be an everywhere 
umbilical hypersurface, whose extrinsic curvature $K_{ij}$ is 
proportional to the induced metric $g_{ij}$, $K_{ij}=\sigma g_{ij}$. 
In general, this condition is weaker than that imposed on a vacuum 
brane, because $\sigma$ of an everywhere umbilical hypersurface need  
not be a constant. However, in a constant curvature space(-time), 
these two conditions become equivalent. This is an immediate 
consequence of the contracted Codazzi equation,
\Eq{
0=n^\mu \tilde R_{\mu ji}{}^j=\mp(n-1)\nabla_i \sigma, 
}
where $n^\mu$ is the unit vector normal to $\Sigma$ with norm $\pm1$,
and $n$ is the dimension of $\tilde M_\lambda$. Hence, we can assume
that $\sigma$ is constant. Further, from the Gauss equation
\Eq{
R_{ijkl}=\tilde R_{ijkl}+K_{ik}K_{jl}-K_{il}K_{jk}
=(\lambda + \sigma^2)(g_{ik}g_{jl}-g_{il}g_{jk}),
}
it follows that $\Sigma$ has constant sectional curvature 
$\lambda+\sigma^2$. Hence, when $\lambda$ is given, $\sigma^2$ 
uniquely determines the intrinsic geometry of $\Sigma$. In this 
section, we show that $\sigma^2$ also uniquely determines the 
configuration of $\Sigma$ up to an isometric transformation of 
$\tilde M$, and thus an everywhere umbilical hypersurface in a constant 
curvature space(-time) is rigid. In this section, we assume that the 
hypersurface is time-like when $\tilde M_\lambda$ is a spacetime.

Let us consider a Gaussian normal coordinate system $(v,x^i)$ in terms 
of which the metric is written
\Eq{
d\tilde s^2=dv^2+ g_{ij}(v,\bm{x})dx^i dx^j,
}
and $\Sigma$ is expressed as $v=v_0$ (constant). Then, the 
definition of $K_{ij}$ and the decomposition of the curvature 
$\tilde R_{vivj}$ of $\tilde M$ yield
\Eqrsubl{GaussianNormal}{
&& \partial_v g_{ij}=2K_{ij},\\
&& \partial_v K_{ij}-K_{il}K^l_j=-\tilde R_{vivj}=-\lambda g_{ij}.
}
With the ansatz $K_{ij}=k(v)g_{ij}$, these equations reduce to
\Eqrsubl{IsotropicSlicing}{
&& \partial_v k + k^2=-\lambda,\\
&& \partial_v g_{ij}=2kg_{ij}.
}
The general solution to the first equation is given by
\Eqrsubl{LIH:k}{
& \lambda=0: & k=\frac{1}{v},\ 0,\\
& \lambda=\frac{1}{\ell^2}: 
& k=\frac{1}{\ell}\cot \frac{v}{\ell},\\
& \lambda=-\frac{1}{\ell^2}: 
& k=\frac{1}{\ell}\coth \frac{v}{\ell},\ \frac{1}{\ell}, \ 
\frac{1}{\ell}\tanh \frac{v}{\ell}.
}
For each $k(v)$, the second equation determines $g_{ij}$  
to be 
\Eq{
g_{ij}=\ell^2\rho(v/\ell)^2\hat g_{ij}(\bm{x}),
\label{ConstCurvature:space}}
where
\Eqrsubl{LIH:rho}{
& \lambda=0: & \rho(\chi)=\chi,\ 1,\\
& \lambda=\frac{1}{\ell^2}: 
& \rho(\chi)=\sin\chi,\\
& \lambda=-\frac{1}{\ell^2}: 
& \rho(\chi)=\sinh \chi,\ e^\chi,\ \cosh\chi.
}
Since \eqref{GaussianNormal} is a set of ordinary differential 
equations with respect to the independent variable $v$ for a fixed value of 
$x^i$, its solution is uniquely determined by $K_{ij}(v_0,\bm{x})$ 
and $g_{ij}(v_0,\bm{x})$. However, because they are related by 
$K_{ij}=\sigma g_{ij}$ at $v=v_0$, and because $k$ given above can be
always made to  
coincide with a given $\sigma$ by choosing $v_0$ appropriately, 
\eqref{ConstCurvature:space} obtained with the ansatz 
$K_{ij}=k(v)g_{ij}$ actually yields the unique global solution to
\eqref{GaussianNormal}. Further, the Gauss equation guarantees that 
$\hat g_{ij}$ represents a normalized constant curvature metric on 
$\Sigma$:
\Eqr{
&& \hat R_{ijkl}=K(\hat g_{ik}\hat g_{jl}-\hat g_{il}\hat g_{jk});\\
&& K\equiv \ell^2\rho(v/\ell)^2(\lambda+k^2)
=\left\{\begin{array}{cccl}
1, & 0 &  & : \lambda=0,\\
1  &   &  & : \lambda=1/\ell^2,\\
1, & 0,& -1 & : \lambda=-1/\ell^2.
\end{array}\right.
}
Hence, the metric can be written
\Eq{
d\tilde s^2=dv^2 + \ell^2\rho(v/\ell)^2 d\sigma_K^2.
\label{CCS:metric:1}}
If $\rho(\chi)\not\equiv1$, by introducing the coordinate $r$ as
$r=\ell\rho(v/\ell)$, this metric can be put into the form
\Eq{
d\tilde s^2=\frac{dr^2}{K-\lambda r^2} + r^2 
d\sigma_K^2,
\label{CCS:metric:2}}
for which the location of the brane, $r=r_0$, and the brane tension 
$\sigma$ are related by
\Eq{
\frac{K}{r_0^2}=\sigma^2+\lambda.
}
Also, for $\rho(\chi)\equiv1$, i.e., for $\lambda=K=0$ 
and $\sigma=0$, the metric can be written 
\Eq{
d\tilde s^2= dv^2+\eta_{ij}dx^idx^j,
\label{CCS:metric:3}}
where $\eta_{ij}$ is a Kronecker delta or a Minkowski 
metric.

Note that the coordinate system in which the metric of $\tilde
M_\lambda$ is written in the form \eqref{CCS:metric:1} and $\Sigma$ is
represented as $v=v_0$ is unique if $\sigma^2$ and $\lambda $ are
given, except for the freedom corresponding to $v/\ell \tend
\pi-v/\ell$ for $\lambda >0$ and the freedom corresponding to
transformations of the form $v \tend v+$const and an associated
rescaling of the flat metric for $K=0$. This implies that two locally
isotropic hypersurfaces with the same value of $\sigma^2$ are mapped
into each other by an isometry of $\tilde M$.

\subsection{Type IV solutions} \label{sec:TypeIV}

From the general argument in the previous subsection, it is clear that
for the type IV solutions, for which the bulk spacetime is a constant
curvature spacetime $M_\lambda^D$, a brane $\Sigma^{D-1}$ with
$K_{\mu\nu}=\sigma g_{\mu\nu}$ is a $(D-1)$-dimensional constant
curvature spacetime $M_K^{D-1}$ with
$K=K_\Sigma\equiv\lambda+\sigma^2$, and its configuration is uniquely
specified by the value of $\sigma^2$ up to isometries of the bulk
spacetime. In this section, we give a general expression for this
brane configuration in a coordinate system in which isometry
transformations have the simplest representation. Then, using this
general expression, we show that the type IV is actually a subclass of
the type I.

\subsubsection{$M^D=E^{D-1,1}$}

For $\lambda=0$, the bulk spacetime is $E^{D-1,1}$, 
and its metric is given by
\Eq{
E^{D-1,1}:\ ds_D^2=-dT^2+d\bm{x}^2.
\label{E:metric}}
The isometry group is represented by the standard Poincare 
group $IO(D-1,1)$ in this coordinate system $(T,x^a)$. 
Since $K_\Sigma=\sigma^2\ge0$, this spacetime can contain two types 
of branes. 

First, for $\sigma=0$, $K_\Sigma=0$ and the brane 
$\Sigma^{D-1}$ is isometric to $E^{D-2,1}$.  Its 
configuration is given by 
\Eq{
\Sigma^{D-1}\cong E^{D-2,1}: qT+\bm{p}\cdot\bm{x}=A;\ 
\bm{p^2}>q^2,
\label{IV:E:sigma=0}}
which can always be put into the form $x^{D-1}=0$ by an isometry. If
we set $R=x^{D-1}$ and introduce the spherical coordinates
$(\chi,\Omega_{D-3}^i)$ for $(x^1,\cdots,x^{D-2})$, the metric
\eqref{E:metric} can be written in the form
\eqref{bulkgeometry:symmetric} with $U=W=S=1$ and $K=1$. In this
coordinate system, the brane configuration $x^{D-1}=A$ is written
$R=A$ and belongs to the type I-C. On the other hand, if we transform
the brane configuration to $x^{D-2}=A$ by an isometry, it can be
written $\chi\cos\theta=A$, which belongs to the type I-A. Further, if
we transform the configuration to $qT+|\bm{p}|X^{D-1}=A$, it belongs
to the type I-B. Thus, every brane configuration with $\sigma=0$ can
be put into a form belonging to the type I, and all solutions with
$\sigma=0$ of the type I are mapped into each other by isometries.

Next, for $\sigma\not=0$, the curvature $K_\Sigma$ of the brane is 
always positive and $\Sigma^{D-1}$ is isometric to 
$\dS^{D-1}(1/|\sigma|)$. Its configuration is given by
\Eq{
\Sigma^{D-1}\cong \dS^{D-1}(1/|\sigma|):\ 
-(T-T_0)^2+(\bm{x}-\bm{a})^2=1/\sigma^2.
\label{IV:E:sigma/=0}}
In the spherical coordinates $(R,\Omega_{D-2}^i)$ for $\bm{x}$, the 
metric \eqref{E:metric} takes the form 
\eqref{bulkgeometry:symmetric} with $U=W=1$, $S=R$ and $K=1$. In 
this coordinate system, the configuration \eqref{IV:E:sigma/=0} with 
$\bm{a}\not=0$ can be written in the form 
\eqref{Sol:R_i/=0:R_T/=0:K/=0:lambda=0} by choosing the coordinate 
$\chi$ as $\bm{x}\cdot\bm{a}=R|\bm{a}|\cos\chi$,  while it belongs 
to the type I-B for $\bm{a}=0$. Hence, \eqref{IV:E:sigma/=0} is also 
isometric to a solution of the type I.

\subsubsection{$M^D=\dS^D$}

For $\lambda=1/\ell^2>0$, the bulk spacetime is a de 
Sitter spacetime $\dS^D(\ell)$, and its metric is expressed in terms 
of a homogeneous coordinate system $Y$ as 
\Eq{
\dS^D:\ ds_D^2=\ell^2 dY\cdot dY;\ 
Y\cdot Y\equiv -(Y^0)^2+(Y^1)^2+\cdots +(Y^{D})^2=1.
\label{dS^D:metric}}
The isometry group is represented as the linear 
transformation group $\OG(D,1)$ in the $Y$-space. Since 
$K_\Sigma=\sigma^2+1/\ell^2>0$, $\Sigma^{D-1}$ is isometric to a 
de Sitter spacetime 
$\dS^{D-1}(\ell/\sqrt{1+\ell^2\sigma^2})$. If we define a 
new coordinate system $(r,\bar Y^M)$ by
\Eq{
r=\ell \sqrt{1-(Y^D)^2},\ 
\bar Y^M=\frac{Y^M}{\sqrt{1-(Y^D)^2}}, \ (M=0,\cdots,D-1)
}
the metric \eqref{dS^D:metric} takes the form of  
\eqref{CCS:metric:2}, with $d\sigma_K^2$ given by
\Eq{
ds_{D-1}^2=d\bar Y\cdot d\bar Y;\
\bar Y\cdot \bar Y\equiv -(\bar Y^0)^2+(\bar Y^1)^2+\cdots 
+(\bar Y^{D-1})^2=1.
\label{dS^(D-1):metric}}
The brane configuration is isometric to the hypersurface 
$r=\ell/\sqrt{1+\ell^2\sigma^2}$ in this coordinate system, and to 
$Y^D=\pm\ell\sigma/\sqrt{1+\ell^2\sigma^2}$ in the $Y$-coordinate 
system. Hence, the  general configuration of a brane is represented as
\Eq{
\Sigma^{D-1}\cong \dS^{D-1}
\left(\frac{\ell}{\sqrt{1+\ell^2\sigma^2}}\right):\ 
P\cdot Y=\frac{\ell\sigma}{\sqrt{1+\ell^2\sigma^2}};\ 
P\cdot P=1.
\label{IV:dS}}

The metric \eqref{dS^D:metric} can be put into the form 
\eqref{bulkgeometry:symmetric} with $U=W=1-R^2/\ell^2$, $S=R$ and 
$K=1$ in the coordinates $(T,R,\Omega_{D-2}^a)$ defined by
\Eq{
Y^0=\sqrt{1-\frac{R^2}{\ell^2}}\sinh\frac{T}{\ell},\ 
Y^D=\sqrt{1-\frac{R^2}{\ell^2}}\cosh\frac{T}{\ell},\ 
Y^a=\frac{R}{\ell}\Omega^a.\ (a=1,\cdots,D-1)
}
Since $X=p_a\Omega^a$ for any constant vector $p_a$ satisfies 
\eqref{eq:X:K/=0} with $K=1$ and $c=p^ap_a$, as shown in Appendix 
\ref{appendix:DDu}, it is easy to see that \eqref{IV:dS} can be  written 
in the form \eqref{Sol:R_i/=0:R_T/=0:K/=0:lambda/=0} when $P^0$ or
$P^D$ is not zero and $(P^a)\not=0$, while it takes the type I-B form
when $P^a=0$. Further, it belongs to the type I-A when $\sigma=0$ and
$P^0=P^D=0$. There exists no representation of the type I-C.

\subsubsection{$M^D=\AdS^D$}

Finally, for $\lambda=-1/\ell^2<0$, the bulk spacetime is 
an anti-de Sitter spacetime $\AdS^D(\ell)$, and its metric 
is expressed in terms of a homogeneous coordinate system $Z$ as
\Eq{
\AdS^D:\ ds_D^2=\ell^2 dZ\cdot dZ;\ 
Z\cdot Z\equiv -(Z^0)^2-(Z^D)^2+(Z^1)^2+\cdots 
+(Z^{D-1})^2=-1.
\label{AdS^D:metric}}
The isometry group is represented as the linear 
transformation group $\OG(D-1,2)$ in the $Z$-space. In 
this case, the curvature $K_\Sigma=\sigma^2-1/\ell^2$ of the brane 
can take any value. 

First, for $\sigma^2\ell^2=1$, $K_\Sigma=0$ and the brane is 
isometric to $E^{D-2,1}$. In the coordinates $(r,x^a)$ 
defined by
\Eq{
r=\ell |Z^D-Z^{D-1}|,\ 
x^a=\frac{Z^a}{Z^D-Z^{D-1}},\ (a=0,\cdots,D-2)
}
\eqref{AdS^D:metric} takes the form of  
\eqref{CCS:metric:2} with $d\sigma_K^2=d\bm{x}\cdot 
d\bm{x}$. Hence, the general configuration of a brane is expressed as
\Eq{
\Sigma^{D-1}\cong E^{D-2,1}: P\cdot Z=A;\ 
P\cdot P=0,
\label{IV:AdS:K=0}}
where $A$ is an arbitrary non-vanishing constant, which can be set  
to unity by a boost in the $Z^D-Z^{D-1}$ plane. If we introduce 
the coordinates $(T,R,z^i)$ by
\Eq{
R=r,\ T-T_0=\frac{\ell^2 Z^0}{r},\ 
z^i=\frac{\ell Z^i}{r}\ (i=1,\cdots,D-2),
}
the metric \eqref{AdS^D:metric} takes the form of  
\eqref{bulkgeometry:symmetric} with $U=W=R^2/\ell^2$, $S=R$ and 
$K=0$. In this coordinate system, the configuration 
\eqref{IV:AdS:K=0} gives the solution 
\eqref{Sol:R_i/=0:R_T/=0:K=0:alpha/=0} if $P^D\not=P^{D-1}$, while 
it gives the solution \eqref{Sol:R_i/=0:R_T/=0:K=0:alpha=0} if
$P^D=P^{D-1}$ and $(P^\mu)\not=0$ ($\mu=0,\cdots,D-2$). Also, for
$P^D=P^{D-1}$ and $P^\mu=0$, the brane configuration becomes $R=$const
and belongs to the type I-C. There exists no representation of the
type I-B.

Second, for $\sigma^2\ell^2>1$, $K_\Sigma>0$ and the brane is 
isometric to $\dS^{D-1}(\ell/\sqrt{\ell^2\sigma^2-1})$. 
In this case, in the coordinate system $(r,\bar Y^M)$ defined  by
\Eq{
r=\ell\sqrt{(Z^D)^2-1},\ 
\bar Y^M=\frac{Z^M}{\sqrt{(Z^D)^2-1}},\ (M=0,\cdots,D-1)
}
the metric \eqref{AdS^D:metric} takes the form of  
\eqref{CCS:metric:2} with $d\sigma_K^2$ given by 
\eqref{dS^(D-1):metric}. Hence, the 
brane configuration is 
\Eq{
\Sigma^{D-1}\cong \dS^{D-1}
\left(\frac{\ell}{\sqrt{\ell^2\sigma^2-1}}\right):\ 
P\cdot Z=\frac{\ell\sigma}{\sqrt{\ell^2\sigma^2-1}};\ 
P\cdot P=-1.
\label{IV:AdS:K=1}}
In the coordinate system $(T,R,\Omega^a)$ defined by
\Eq{
Z^0=\sqrt{1+\frac{R^2}{\ell^2}}\sin\frac{T}{\ell},\ 
Z^D=\sqrt{1+\frac{R^2}{\ell^2}}\cos\frac{T}{\ell},\ 
Z^a=\frac{R}{\ell}\Omega^a,\ (a=1,\cdots,D-1)
}
the metric \eqref{AdS^D:metric} can be written in the form of 
\eqref{bulkgeometry:symmetric} with $U=W=1+R^2/\ell^2$, $S=R$ and 
$K=1$. When $(P^a)$ does not vanish, the brane configuration 
\eqref{IV:AdS:K=1} in this coordinate system gives 
\eqref{Sol:R_i/=0:R_T/=0:K/=0:lambda/=0} by setting $X=P_a\Omega^a$, 
while for $(P^a)=0$, it gives the type I-B solution. There exists no 
representation of the type I-C.

Finally, for $\sigma^2\ell^2<1$, $K_\Sigma<0$ and the brane is 
isometric to $\AdS^{D-1}(\ell/\sqrt{1-\ell^2\sigma^2})$. 
In the coordinate system $(r,\bar Z^M)$ defined by
\Eq{
r=\ell\sqrt{(Z^{D-1})^2+1},\ 
\bar Z^M=\frac{Z^M}{\sqrt{(Z^{D-1})^2+1}},\ 
(M=0,\cdots,D-2,D)
}
the metric \eqref{AdS^D:metric} takes the form of  
\eqref{CCS:metric:2} with $d\sigma_K^2$ given by
\Eq{
ds_{D-1}^2=d\bar Z\cdot d\bar Z;\ 
\bar Z\cdot\bar Z\equiv -(\bar Z^0)^2-(\bar Z^D)^2+
(\bar Z^1)^2+\cdots (\bar Z^{D-2})^2=-1.
}
Hence, the brane configuration is 
given by
\Eq{
\Sigma^{D-1}\cong \AdS^{D-1}
\left(\frac{\ell}{\sqrt{1-\ell^2\sigma^2}}\right):\ 
P\cdot Z=\frac{\ell\sigma}{\sqrt{1-\ell^2\sigma^2}}; \ 
P\cdot P=1.
}
In the coordinate system $(T,R,\bar Y^a)$ defined by
\Eqr{
&& Z^0=\sqrt{-1+\frac{R^2}{\ell^2}}\sinh\frac{T}{\ell},\ 
Z^{D-1}=\sqrt{-1+\frac{R^2}{\ell^2}}\cosh\frac{T}{\ell},\nonumber\\ 
&& Z^a=\frac{R}{\ell}\bar Y^a,\ (a=1,\cdots,D-2,D)
}
the metric \eqref{AdS^D:metric} takes the form of  
\eqref{bulkgeometry:symmetric} with $U=W=-1+R^2/\ell^2$, $S=R$ and 
$K=-1$. Since $X=p_a\bar Y^a$ for any constant vector $p_a$ 
satisfies \eqref{eq:X:K/=0} with $K=-1$ and $c=p^ap_a$, as shown in 
Appendix \ref{appendix:DDu}, the brane configuration 
\eqref{IV:AdS:K=1} in this coordinate system gives 
\eqref{Sol:R_i/=0:R_T/=0:K/=0:lambda/=0} when $(P^a)$ does not 
vanish, while for $(P^a)=0$ and $\sigma\not=0$ it gives the type I-B 
solution. Further, for $P^0=P^{D-1}=0$ and $\sigma=0$, it can be 
written in the type I-A form. There exists no representation of the 
type I-C for $\sigma\not=0$ and of the type I-B or the  type I-C for 
$\sigma=0$.

\begin{table}[t]
\caption{\label{table:conf:TypeIV}Type I configurations in the type 
IV geometry.}
\begin{center}
\begin{tabular}{lll}
\hline\hline
Spacetime  &  $\sigma$ & Type I configurations \\
\hline
&&\\
$E^{D-1,1}$: 
& $\sigma=0$   & I-A,\ I-B,\ I-C\ (mutually isometric) \\
& $\sigma\not=0$ & I-B \\
&&\\
$\dS^D$:
& $\sigma=0$   & I-A,\ I-B (mutually isometric) \\
& $\sigma\not=0$ & I-B \\
&&\\
$\AdS^D$:
& $\sigma=0$   & I-A \\
& $\sigma\not=0$& I-B ($\ell^2\sigma^2\not=1$) \ 
                  I-C ($\ell^2\sigma^2=1$) \\
&&\\
\hline
\hline
\end{tabular}
\end{center}
\end{table}

\subsubsection{Summary of the type IV solutions}

The results obtained in this subsection can be summarized as 
follows. First, for a brane with $\sigma=0$, solutions are mutually 
isometric and always have representations of the type I-A. They also 
have type I-B representations, except for $M^D=\AdS^D$, but a type 
I-C representation exists only for $M^D=E^{D-1,1}$. Second, for a 
brane with $\sigma\not=0$, solutions with the same $\sigma^2$ are 
mutually isometric and have only representations of the type I-B, 
except for the brane with $\ell^2\sigma^2=1$ in $\AdS^D$. In this 
exceptional case, there exist only type I-C representations. In 
order to make the comparison of these features with those for the 
other types easier, we list the possible type I representations for 
each constant curvature spacetime in Table \ref{table:conf:TypeIV}.

Note also that every brane configuration is invariant under some 
group isomorphic to $\IO(D-2,1)$, $\OG(D-1,1)$ or $\OG(D-2,2)$. In 
particular, they are static, but not invariant under $\IO(1)\times 
G(D-2,K)$, except in the $\IO(D-2,1)$-invariant case. 

\subsection{Type III solutions}

The type III geometry can be written 
\Eq{
ds_D^2=-R^2dT^2+\frac{dR^2}{W} + R^2d\bm{z}^2,
}
where $W$ is not proportional to $R^2$. Then, from the arguments in 
\S\ref{sec:Sol:S=R}, the following four families of solutions can exist. 
The first family represents a brane with $\sigma=0$ and configuration
$R=$const. This is of the type I-C and exists only when $f'(v)=0$ has
a solution. The second family represents a brane with configuration
$R=R(\bm{z})$ with $R_i\not\equiv0$, which belongs to the type
II. Since the type II and the type III are mutually exclusive, as
shown below, this case cannot be realized. The third family represents
a brane with $\sigma=0$ belonging to the type I-A, which exists for
any bulk geometry. The final family is a solution with $\sigma=0$ and
configuration $\chi=\chi(T,\bm{z})$ with $\chi_T\not\equiv0$. It must
correspond to a time-like hyperplane in the $x$-space $E^{D-2,1}$, and
can always be expressed in the form I-A corresponding to the third
family, in an appropriate coordinate system.

We thus find that a bulk spacetime of the type III allows only branes
with $\sigma=0$. Branes can have two geometrically different types of
configurations. The first one is of the type I-A, which always exists. 
Configurations of this type are mutually isometric and do not have a
I-B type representation, as shown below. The second one is of the type
I-C, which is allowed only when $f'(v)=0$ has a solution. Solutions of
this type are not isometric to those of the type I-A.

\subsection{Type II solutions}\label{sec:TypeII}

The bulk geometry of this type is described by 
\eqref{TypeII:metric} and has the structure 
$(E^{D-1},E^{0,1})$, $(S^{D-1},E^{0,1})$ or $(H^{D-1},E^{0,1})$,
depending on the sign of $\lambda$. Since this type is mutually
exclusive with the types III and IV, as shown below, from the
arguments at the beginning of this section, a solution belonging to
this type must have a type I-B representation if it is not
static. However, the bulk geometry is of the type IV if
$U=W$. Further, only $\IO(D-2)$-invariant branes with $\sigma=0$ in
$(E^{D-1},E^{0,1})$ can have non-static configurations of the type
I-B, but the bulk metric must take the form
\eqref{Sol:Naria:R_i/=0:metric} in this case, which is excluded in the
definition of the type II. Therefore, all the type II solutions are
static, and are represented as $F(x)=0$ in terms of a function on the
base space $M_\lambda^{D-1}$. This implies that the brane projects to
an everywhere umbilical hypersurface with $K_{ab}=\sigma g_{ab}$ in
the base space. In this section, using this fact, we show that
brane configurations in type II geometries belong to the type I,
unless $U(R)$ takes some special form, and classify possible
configurations instead of using the explicit solutions obtained in
\S\ref{sec:Sol:S=R}, because it is much simpler. The main
result is summarized in Table \ref{table:conf:TypeII}.

\begin{table}[t]
\caption{\label{table:conf:TypeII}Type I configurations in the type 
II geometry.}
\begin{center}
\begin{tabular}{lll}
\hline\hline
Spacetime  &  $\sigma$ & Type I configurations \\
\hline
&&\\
$(E^{D-1},E^{0,1})$: 
& $\sigma=0$   & I-A \\
& $\sigma\not=0$& I-C\ ($U\not\propto((\bm{x}-\bm{a})^2+k)^2$) \\
&&\\
$(S^{D-1},E^{0,1})$:
& $\sigma=0$   & I-A\ (\& I-C) \\
& $\sigma\not=0$& I-C ($U\not\propto(P\cdot X+k)^2$)\\
&&\\
$(H^{D-1}, E^{0,1})$:
& $\sigma=0$   & I-A,\ (\& I-C for $K=-1$) \\
& $\sigma\not=0$& I-C ($U\not\propto(P\cdot Y+k)^2$)\\
&&\\
\hline
\hline
\end{tabular}
\end{center}
\end{table}

The general scheme of the argument is as follows. First, 
we relate a coordinate system in which the metric of 
$M_\lambda^{D-1}$ is written in the decomposed form 
\eqref{CCS:metric:2} or \eqref{CCS:metric:3}  and a global 
homogeneous coordinate system for 
$M_\lambda^{D-1}$ in which the action of the isometry 
group has a simple representation, as was done in 
\S\ref{sec:TypeIV}. Then, since both the argument of 
$U$ and the function $F(x)$ defining the brane 
$\Sigma^{D-1}$ are obtained from a $v$-coordinate giving 
the decomposition \eqref{CCS:metric:1} through 
isometries of $M_\lambda^{D-1}$, we can easily find 
possible canonical forms for the argument of $U$ and 
$F(x)$. Next, we determine $U$ by the condition 
$K_{TT}=\sigma g_{TT}$, which is simply written 
\Eq{
\pm 
\frac{\nabla F\cdot \nabla U}{2U|\nabla F|}=\sigma
\label{TypeII:K_TT}}
in the present case, where the sign on the left-hand side 
must be chosen so that the normal vector $\pm \nabla 
F/|\nabla F|$ gives the correct sign in  $K_{ab}=\sigma g_{ab}$.

\subsubsection{$M^D=(E^{D-1},E^{0,1})$} 

In this case, the bulk metric is written
\Eq{
ds_D^2=d\bm{x}^2 -UdT^2.
\label{TypeII:lambda=0:metric}}
Since $\lambda=0$, the decomposition \eqref{CCS:metric:1} 
gives $K=0$ or $K=1$, for which $v$ in \eqref{CCS:metric:3} or $r$ 
in \eqref{CCS:metric:2} depends only on $x^{D-1}$ or $\bm{x}^2$, 
respectively,  after some isometric transformations. Hence, we can 
assume that $U$ 
is a function of $x^{D-1}$ or $\bm{x}^2$, and the generic 
form of a static brane configuration is 
\Eqrsubl{TypeII:E:brane}{
& \sigma=0: & F(\bm{x})\equiv\bm{p}\cdot\bm{x}-A=0,
\label{TypeII:E:brane:sigma=0}\\
& \sigma\not=0: & F(\bm{x})\equiv(\bm{x}-\bm{a})^2-1/\sigma^2=0,
\label{TypeII:E:brane:sigma/=0}
}
where $\bm{p}$ and $\bm{a}$ are $(D-1)$-dimensional constant vectors 
and $A$ is a constant.

We first consider the case $U=U(x^{D-1})$. In this case, 
for a brane with $\sigma=0$, $F$ can be put into the form 
$F=qx^{D-2}+p x^{D-1}-A$ by an isometry of $E^{D-1}$ 
preserving $U$. Then, \eqref{TypeII:K_TT} reduces to 
$pU'(x^{D-1})=0$. If $p\not=0$, we obtain 
$U'=0$. Hence, if $p\not=0$ and $q\not=0$, $U$ must 
be constant, and the bulk metric 
\eqref{TypeII:lambda=0:metric} represents a flat metric, 
which implies that the solution belongs to the type IV. On 
the other hand, if $p\not=0$ and $q=0$, $U$ can be 
arbitrary, and the brane configuration is given by a 
solution to $U'(x^{D-1})=0$. Hence, the solution belongs 
to the type I-C. $U$ can be arbitrary as well for $p=0$, 
but in this case, the brane configuration is isometric to 
$x^{D-2}=0$. It is easy to see that this solution belongs 
to the type I-A when expressed in terms of appropriate spherical 
coordinates for $E^{D-1}$. 

Next, for $\sigma\not=0$, \eqref{TypeII:K_TT} with the brane 
configuration \eqref{TypeII:E:brane:sigma/=0}  
becomes $U'/U=2(x^{D-1}-a^{D-1})$. Hence, $U$ can be put into 
the form $U=(x^{D-1})^2$ by an isometry and a constant 
rescaling of $T$. Then, the bulk metric takes the Rindler 
form of the flat metric, and the solution belongs to the 
type IV.

These results for $U=U(x^{D-1})$ justify the exclusion of the bulk 
geometry represented as $ds_D^2=-U(R)dT^2+dR^2+d\bm{z}^2$ from the 
type II.

For the case $U=U(\bm{x}^2)$, \eqref{TypeII:K_TT} for a brane with 
$\sigma=0$ is equivalent to $AU'=0$ on the brane. Hence, when $A=0$, 
$U$ is arbitrary, and the brane configuration is isometric to 
$x^{D-2}=x^{D-1}$. This solution belongs to the type I-A. On the other 
hand, when $A\not=0$, we obtain the constraint 
$U=$const for $\bm{p}^2\bm{x}^2\ge A^2$. If $U$ is an analytic 
function, this solution belongs to the type IV. 

Next, for a brane with $\sigma\not=0$,  \eqref{TypeII:K_TT} with 
\eqref{TypeII:E:brane:sigma/=0} reads
\Eq{
\frac{U'}{U}=\frac{2}{1/\sigma^2-\bm{a}^2+\bm{x}^2}.
\label{TypeII:K_TT:E:K=1}}
For $\bm{a}\not=0$, this determines $U$ to be proportional to 
\eqref{Sol:R_i/=0:R_T=0:K/=0:lambda=0:U} with $R=|\bm{x}|$, and the 
brane configuration cannot have a type I representation. In 
contrast, for $\bm{a}=0$, this equation gives an equation for 
$\sigma^2$, and its solution (if one exists) belongs to the type I-C.

To summarize, if the metric \eqref{TypeII:lambda=0:metric} does  not 
have a constant curvature and $U$ is not of the form 
$U(\bm{p}\cdot\bm{x})$, a brane with $\sigma=0$ always has static 
configurations of the type I-A, which are mutually isometric. If $U$ 
is an analytic function, these exhaust all possible configurations of 
a brane with $\sigma=0$.  Similarly, a brane with $\sigma\not=0$ is 
always static in the type II bulk spacetime. Further, except for 
exceptional configurations of the type I-C for an $O(D-1)$-invariant 
$U$,  a brane with $\sigma\not=0$ can be embedded in the type II 
geometry only if $U$ has the special form %
\Eq{
U=(\bm{x}^2+k)^2.
\label{TypeII:E:U}}
Brane configurations in this geometry are given by 
\eqref{TypeII:E:brane:sigma/=0} with
\Eq{
\bm{a}^2=\frac{1}{\sigma^2}-k,
}
and do not have a type I representation. Here, note that, strictly 
speaking, $U$ has to take this form only in the region 
\Eq{
(|\bm{a}|-1/|\sigma|)^2\le \bm{x}^2 \le (|\bm{a}|+1/|\sigma|)^2,
}
unless the analyticity of $U$ is assumed.

\subsubsection{$M^D=(S^{D-1},E^{0,1})$}\label{sec:TypeII:S}

In this case, the bulk metric is written
\Eq{
ds_D^2=\ell^2dX\cdot dX -UdT^2;\ 
X\cdot X\equiv (X^1)^2+\cdots+(X^D)^2=1.
\label{TypeII:lambda>0:metric}}
Since $\lambda>0$, the decomposition \eqref{CCS:metric:1} 
gives $K=1$, and $v=$const surfaces are given by the 
intersections of parallel hyperplanes and the sphere $X\cdot X=1$ in 
the $X$-space. Hence, with an appropriate choice of the 
$X$-coordinates, we can assume that $U$ depends only on $X^D$ and 
the brane is represented as 
\Eq{
P\cdot X=\ell\sigma;\ 
P\cdot P=1+\ell^2\sigma^2.
\label{TypeII:lambda>0:brane}}
For this choice, since $X^M$ ($M=1,\cdots,D$) satisfies
\Eq{
\nabla X^M\cdot \nabla X^N=\delta^{MN}-X^M X^N,
\label{Id:S}}
where $\nabla$ is the covariant derivative of the unit 
$(D-1)$-dimensional sphere, \eqref{TypeII:K_TT} gives
\Eq{
(P^D -\ell\sigma X^D)\frac{U'}{U}=-2\ell\sigma.
\label{TypeII:K_TT:S}}

First, for a brane with $\sigma=0$, this equation becomes trivial 
for configurations with $P^D=0$, which belongs to the type I-A. In 
contrast, for $P^D\not=0$ and $P^a=0$ ($a=1,\cdots,D-1$), this 
equation reads $U'(0)=0$, and if this condition is satisfied, there 
exists a configuration of the type I-C. Finally, configurations with 
$P^D\not=0$ and $(P^a)\not=0$ are allowed only when $U$ is constant. 
This is the only special case in which $(S^{D-1},E^{0,1})$ has an 
extra continuous symmetry, as shown in Table 
\ref{table:SymmetryList}. In this special geometry, configurations 
of the type I-C are isometric to those of the type I-A. 

Next, for a brane with $\sigma\not=0$ and $P^a=0$, 
\eqref{TypeII:K_TT:S} is equivalent to the equation for the 
location of a type I-C brane, while for a brane with $\sigma\not=0$ 
and $(P^a)\not=0$,  it requires  $U$ to be proportional to 
$(P^D-\ell\sigma X^D)^2$. Hence, the bulk spacetime with the  
structure $(S^{D-1},E^{0,1})$ can contain a brane with 
$\sigma\not=0$ that is not of the type I-C only when $U$ has the 
form
\Eq{
U=(X^D+k)^2,
\label{TypeII:S:U}}
where $k\not=0$, because the spacetime becomes $\dS^D$ for  
$U=(X^D)^2$. The corresponding brane configuration is written
\Eq{
-kX^D+ p_a X^a=1;\ k^2+\bm{p}^2=1+\frac{1}{\ell^2\sigma^2},
\label{TypeII:S:brane}}
which does not have a type I representation. It is easy to see that 
this can be expressed in the form of  
\eqref{Sol:R_i/=0:R_T=0:K/=0:lambda/=0} with $K=1$ and 
$\lambda=1/\ell^2$ if we set $R=\ell\sqrt{1-(X^D)^2}$ and $X=\mp 
a\ell\sigma p_a X^a/\sqrt{1-(X^D)^2}$, because $X=b_a\Omega^a$ 
satisfies  \eqref{eq:X:K/=0} with $K=1$ and $c=b_ab^a$ on the unit 
sphere $\Omega_a\Omega^a=1$, as shown in Appendix \ref{appendix:DDu}.

\subsubsection{$M^D=(H^{D-1},E^{0,1})$}

In this case, the bulk metric is written
\Eq{
ds_D^2=\ell^2dY\cdot dY -UdT^2;\ 
Y\cdot Y\equiv -(Y^0)^2+(Y^1)^2+\cdots+(Y^{D-1})^2=-1,\ Y^0\ge1.
\label{TypeII:lambda<0:metric}}
Now, the hyperbolic space $H^{D-1}$ has tree types of slicing by 
constant curvature hypersurfaces: The decomposition of the 
form \eqref{CCS:metric:2} with $K=0,+1$ and $-1$ is obtained for 
$r=\ell (Y^0-Y^{D-1}), \ell\sqrt{(Y^0)^2-1}$ and $\ell 
\sqrt{(Y^{D-1})^2+1}$, respectively. Since a generic decomposition 
is isometric to one of these, the level surfaces of $r$ are always 
given by the 
intersections of hyperplanes $P\cdot Y=$const with the 
hyperboloid $Y\cdot Y=-1$ in the $Y$-space, and the sign of $K$ 
is the same as that of $-P\cdot P$. In particular, the generic 
configuration of a brane is expressed as
\Eq{
P\cdot Y=A;\quad P\cdot P=1-\ell^2\sigma^2,
\label{TypeII:H:brane}}
where 
\Eq{
A=\ell\sigma \ (\ell^2\sigma^2\not=1), \quad
A\not=0\ (\ell^2\sigma^2=1).
\label{TypeII:H:brane:constraint}}
%

Let us first treat the case $U=U(Y^{D-1})$. Since $Y^M$ 
($M=0,\cdots,D-1$) as a function on $H^{D-1}$ satisfies
\Eq{
\nabla Y^M\cdot\nabla Y^N=\eta^{MN}+Y^M Y^N,
\label{Id:H}}
where $\nabla$ is the covariant derivative for the unit hyperboloid 
$Y\cdot Y=-1$, \eqref{TypeII:K_TT} reads
\Eq{
(P^{D-1}+A Y^{D-1})\frac{U'}{U} 
=2\ell\sigma\sqrt{1-\ell^2\sigma^2+A^2}.
\label{TypeII:K_TT:H:K=-1}}

First, for a brane with $\sigma=0$, $A=0$ from 
\eqref{TypeII:H:brane:constraint} and the left-hand side of this 
equation must vanish. When $P^{D-1}=0$, $U$ can be arbitrary, and 
the brane configuration can be transformed into $Y^{D-3}=Y^{D-2}$. 
This gives a type I-A solution when it is expressed in appropriate 
spherical coordinates for $Y^M$. In contrast, for $P^{D-1}\not=0$ 
and $P^\mu=0$ ($\mu=0,\cdots, D-2$), 
\eqref{TypeII:H:brane:constraint} reduces to $U'(0)=0$, and if this
equation is satisfied, there exists a type I-C solution. Finally,
configurations with $P^{D-1}\not=0$ and $(P^\mu)\not=0$ are possible
only when $U$ is constant. This is the only special case in which the
bulk spacetime $(H^{D-1},E^{0,1})$ has an additional continuous
symmetry, as shown in Table \ref{table:SymmetryList}. In this special
case, all brane configurations with $\sigma=0$ are mutually isometric
and have a type I-A representation.
 
Next, for $\sigma\not=0$, $A$ does not vanish, as seen from 
\eqref{TypeII:H:brane:constraint}, and the right-hand side of 
\eqref{TypeII:K_TT:H:K=-1} is a non-vanishing constant. Hence, if 
$(P^\mu)\not=0$, $U$ must be proportional to $(P^{D-1}+AY^{D-1})^2$, 
and by a constant rescaling of $T$, it can be written
\Eq{
U=(Y^{D-1}+k)^2,
\label{TypeII:H:K=-1:U}}
where $k\not=0$, because the bulk spacetime becomes $\AdS^D$ for 
$U=(Y^{D-1})^2$. The corresponding brane configuration is expressed 
as
\Eq{
kY^{D-1}+p_\mu Y^\mu=1;\quad
k^2+p_\mu p^\mu=\frac{1}{\ell^2\sigma^2}-1,
\label{TypeII:H:K=-1:brane}}
which does not have a type I representation. This corresponds to the 
solution \eqref{Sol:R_i/=0:R_T=0:K/=0:lambda/=0} with $K=-1$ and 
$\lambda=-1/\ell^2$ for $R=\ell\sqrt{(Y^{D-1})^2+1}$ and $X=\pm 
a\ell^2\sigma p_\mu Y^\mu/R$, because $X=b_\mu \bar Y^\mu$ satisfies 
\eqref{eq:X:K/=0} with $K=-1$ and $c=b_\mu b^\mu$ on the unit 
hyperboloid $\bar Y\cdot \bar Y=-1$, as shown in Appendix
\ref{appendix:DDu}. If $k=0$, this solution belongs to the 
type IV. On the other hand, if $P^\mu=0$, we obtain a solution
belonging to the type I-C.

Next, let us consider the case $U=U(Y^0-Y^{D-1})$. In this case, 
\eqref{TypeII:K_TT} reads
\Eq{
[P^0-P^{D-1}+A(Y^0- Y^{D-1})]\frac{U'}{U} 
=2\ell\sigma\sqrt{1-\ell^2\sigma^2+A^2}.
\label{TypeII:K_TT:H:K=0}}
For $\sigma=0$, we can conclude that this equation becomes trivial 
for type I-A configurations, and there exists a type I-C 
configuration if $U'(0)=0$, by the argument given in the 
previous case. For $\sigma\not=0$, if the brane configuration is not 
of the type I-C, \eqref{TypeII:K_TT:H:K=0} constrains $U$ as
\Eq{
U=(Y^0-Y^{D-1}+k)^2,
\label{TypeII:H:K=0:U}}
where $M^D=\AdS^D$ if $k=0$. The corresponding brane configuration 
is expressed as
\Eq{
-k(Y^0+Y^{D-1})+q (Y^0-Y^{D-1})+2p_i Y^i=2;\quad
kq+p_ip^i=\frac{1}{\ell^2\sigma^2}-1,
\label{TypeII:H:K=0:brane}}
where $q$ and $p_i$ ($i=1,\cdots, D-2$) are constants, and does not 
have a type I representation. By a Lorentz transformation of $Y$ and 
a rescaling of $U$, this brane configuration can be put into the 
canonical form
\Eq{
\left(2-\frac{1}{\ell^2\sigma^2}\right)Y^0+\frac{1}{\ell^2\sigma^2}Y^
{D-1}=-2k; \quad k=\pm1,
}
which can be further rewritten in the form 
\eqref{Sol:R_i/=0:R_T=0:K=0:alpha/=0} with $\lambda=-1/\ell^2$, if 
we introduce the coordinates $(R,\bm{z})$ as $R=\ell|Y^0-Y^{D-1}|$ 
and $z^i=Y^i/(Y^0-Y^{D-1})$. 

Finally, we consider the case $U=U(Y^0)$. In this case, 
\eqref{TypeII:K_TT} reads
\Eq{
(P^0+AY^0)\frac{U'}{U} 
=2\ell\sigma\sqrt{1-\ell^2\sigma^2+A^2}.
\label{TypeII:K_TT:H:K=1}}
For $\sigma=0$, we can again conclude that this equation becomes trivial 
for type I-A configurations, and there exists a type I-C 
configuration if $U'(0)=0$. Further, if there exists a brane with  
$\sigma\not=0$ that is not of the type I-C, 
\eqref{TypeII:K_TT:H:K=1} leads to the solution
\Eq{
U=(Y^0+k)^2,
\label{TypeII:H:K=1:U}}
where the spacetime again becomes $\AdS^D$ for $k=0$. The 
corresponding brane configuration is expressed as
\Eq{
-kY^0+p_a Y^a=1;\quad
p_ap^a=k^2-1+\frac{1}{\ell^2\sigma^2},
\label{TypeII:H:K=1:brane}}
where $a$ runs over $1,\cdots, D-1$, and does not have a type I 
representation. This configuration can be put into the form 
\eqref{Sol:R_i/=0:R_T=0:K/=0:lambda/=0} with $\lambda=-1/\ell^2$,  
$K=1$ and $X=\pm a\ell\sigma p_a \Omega^a$ in the coordinates 
$R=\ell\sqrt{(Y^0)^2-1}$ and $\Omega^a=\ell Y^a/R$.


\subsection{Relations among the types}

By the arguments up to this point, it is clear that the types III and 
IV are subclasses of the type I, and solutions of the type II also 
belong to the type I, unless $\sigma\not=0$ and $U$ is proportional 
to \eqref{TypeII:E:U}, \eqref{TypeII:S:U}, 
\eqref{TypeII:H:K=-1:U}, \eqref{TypeII:H:K=0:U} or 
\eqref{TypeII:H:K=1:U} in some homogeneous coordinate system for the 
constant curvature space $M_\lambda^{D-1}$. In this section, we show 
that the type-II solutions with $\sigma\not=0$ for these forms of 
$U$ really do not belong to the type I. We also clarify relations 
among the bulk geometries of the types I-B, II, III and IV.

\subsubsection{The type II vs the types I-B and IV}

We can show that the type II is mutually exclusive with the type IV by 
calculating the Riemann tensor for the metric 
\eqref{TypeII:metric}. First, note that when this metric is put 
into the form \eqref{bulkgeometry:symmetric}, we obtain 
$(K,W,S)=(0,1,1)$ and $(1,1,R)$ for $\lambda=0$, and $W=K-\lambda 
R^2$ and $S=R$ for $\lambda\not=0$.  Inserting these into 
\eqref{Riemann:bulk}, we find 
\Eqr{
&& \tilde R_{ijkl}=\lambda
S^4(\gamma_{ik}\gamma_{jl}-\gamma_{il}\gamma_{jk}),\\ && \tilde
R^i_{RjR}=\frac{\lambda}{W}\delta^i_j,\\ && \tilde
R^i_{TjT}=\frac{WU'S'}{2S}\delta^i_j.  }
Hence, if the spacetime has a constant curvature, $WU'S'/(2US)$ must 
be equal to $-\lambda$. This requires that $U$ be proportional to 
$W$, except in the case $K=0$ and $W=S=1$. For this exceptional 
case, the condition ${}^2R=2\lambda$ requires that $U$ be 
proportional to $(aR+b)^2$, which can be set to $1$ or $R^2$ by a 
constant shift of $R$ and a rescaling of $T$. This result, with the 
argument in \S\ref{sec:TypeII}, shows that the type I-B and the type 
II are also mutually exclusive.

\subsubsection{The type III vs the types I-B, II and IV}

The metric \eqref{TypeIII:metric} corresponds to 
\eqref{bulkgeometry:symmetric} with $R=v$, $U=f^2(v)$, $W=1$, 
$S=f(v)$ and $K=0$. Hence, from \eqref{Riemann:bulk}, we obtain
\Eq{
\tilde 
R_{ijkl}=-f^2(f')^2(\gamma_{ik}\gamma_{jl}-\gamma_{il}\gamma_{jk}).
}
From this, it follows that the spacetime has a constant curvature 
only when $f=$const or $f=f_0e^{v/\ell}$, which correspond to 
$E^{D-1,1}$ and 
$\AdS^D(\ell)$, respectively.

This results also implies that the type III geometry cannot be of the 
type II for the choice of the time coordinate $T$ in terms of which 
the metric is written in the form \eqref{TypeIII:metric}. Hence, if 
we can show that all time-like Killing vectors preserve the metric 
$\eta_{\mu\nu}dx^\mu dx^\nu$ of the fibre $E^{D-2,1}$, the mutual
exclusiveness of the type III and the type II geometries is 
concluded.

Let $\xi=\xi^v\partial_v + \xi^\mu \partial_\mu$ be a 
Killing vector of \eqref{TypeIII:metric}. Then, since the 
non-vanishing components of the Christoffel symbols are 
given by
\Eq{
\Gamma^v_{\mu\nu}=-ff'\eta_{\mu\nu},\ 
\Gamma^\mu_{v\nu}=\frac{f'}{f}\delta^\mu_\nu,
}
the Killing equation yields the three equations
\Eqrsubl{TypeIII:KillingEq}{
&& \partial_v\xi_v=0,\\
&& \partial_v(f^{-2}\xi_\mu)+f^{-2}\partial_\mu\xi_v=0,\\
&& \partial_\mu\xi_\nu + 
\partial_\nu\xi_\mu+2ff'\xi_v\eta_{\mu\nu}=0.
}
The first equation implies that $\xi_v$ depends only on 
$x^\mu$, and the integration of the second equation gives
\Eq{
f^{-2}\xi_\mu=-\partial_\mu \xi_v \int f^{-2}dv+\eta_\mu 
(x),
}
where $\eta_\mu(x)$ is a field that is independent of $v$. Putting 
this into the third equation, we obtain
\Eq{
\partial_\mu\eta_\nu + \partial_\nu\eta_\mu
-2\partial_\mu\partial_\nu \xi_v \int f^{-2}dv
+2\frac{f'}{f}\xi_v\eta_{\mu\nu}=0.
\label{TypeIII:KillingEq:reduced}}

If $1/f^2$ and $(f'/f)'$ are linearly independent, it follows from 
\eqref{TypeIII:KillingEq:reduced} that $\xi_v\equiv0$, which implies 
that the Killing vector preserves the fibre metric. Therefore, let us
assume that $(f'/f)'=a/f^2$, where $a$ is a non-vanishing constant
because the spacetime is not of the type IV. Then, $f$ can be written
$f^2=1/(ar^2+2br+c)$ in terms of the variable $r=\int
dv/f^2$. Inserting this into \eqref{TypeIII:KillingEq:reduced}, we
obtain $\partial_\mu\partial_\nu\xi_v +a\xi_v=0$. It follows from this
equation that $a\partial_\mu\xi_v=0$, and hence $\xi_v=0$. Therefore,
all time-like Killing vectors of the type III geometry preserve the
fibre metric. This result also implies that the type III geometry is
mutually exclusive with the type I-B geometry.

\begin{table}[t]
\caption{\label{table:SymmetryList}A list of geometries with higher 
symmetries}
\begin{tabular}{lll}
\hline\hline
Type & Isometry & Metric \\
\hline
{\bf Type I} &&\\
$E^{1,1}\times M_K^{D-2}$
& $\IO(1,1)\times G(D-2,K)$ 
& $ds^2=-dT^2 + dr^2 + \ell^2 d\sigma_K^2$ \\
$(K=\pm1)$
& ibid
& $ds^2=-r^2dT^2+dr^2+\ell^2 d\sigma_K^2$ \\
$\dS^2(1/\mu)\times M_K^{D-2}$
& $\OG(2,1)\times G(D-2,K)$ 
& $ds^2=-(1-\mu^2 R^2)dT^2+\frac{dR^2}{1-\mu^2R^2}
+\ell^2d\sigma_K^2$ \\
$\AdS^2(1/\mu)\times M_K^{D-2}$
& $\OG(1,2)\times G(D-2,K)$ 
& $ds^2=-(K'+\mu^2 R^2)dT^2+\frac{dR^2}{K'+\mu^2R^2}
+\ell^2d\sigma_K^2$ \\
$(M^{1,1},E^{D-2})$
& $\RF_+\times \IO(1)\times \IO(D-2)$
& $ds^2=-\left(\frac{R}{\ell}\right)^{2n}dT^2
+\frac{\ell^2}{R^2}dR^2+R^2d\bm{x}^2\ (n\not=1)$ \\
&&\\
{\bf Type II} && \\
$E^{0,1}\times M_\lambda^{D-1}$ 
& $\IO(1)\times G(D-1,\lambda)$
& $ds^2=-dT^2+\frac{dR^2}{K-\lambda R^2} + R^2 d\sigma_K^2$; \
($\lambda\not=0$) \\
&&\\
{\bf Type III} && \\
$(E^1,E^{D-2,1})$ 
& $\IO(D-2,1)$
& $ds^2=dv^2+f(v)^2 \eta_{\mu\nu}dx^\mu dx^\nu$;\ 
$\left(\frac{f'}{f}\right)'\not=0$ \\
&&\\
{\bf Type IV} && \\
$E^{D-1,1}$
& $\IO(D-1,1)$
& $ds^2=-dT^2+dr^2 + d\bm{x}^2$,\\
&& $ds^2=-dT^2 + dR^2 + R^2d\Omega_{D-2}^2$,\\
&& $ds^2=-r^2dT^2+dr^2+d\bm{x}^2$ \\
$\dS^D(\ell)$
& $\OG(D,1)$
& $ds^2=-(1-R^2/\ell^2)dT^2+\frac{dR^2}{1-R^2/\ell^2}
+R^2d\Omega_{D-2}^2$\\
$\AdS^D(\ell)$
& $\OG(D-1,2)$
& $ds^2=-(K+R^2/\ell^2)dT^2+\frac{dR^2}{K+R^2/\ell^2}
+R^2d\sigma_K^2$\\
\hline
\hline
\end{tabular}
\end{table}

\subsubsection{Types I-A, I-B and I-C}

We have shown that the bulk geometries of the types I-B, II and III 
as well as the types II, III and IV are never mutually isometric. We 
have also clarified the relations between the type I and the types II, III 
and IV. Therefore, we can complete the geometrical classification of 
solutions if we clarify the uniqueness of the type I representations 
and the relations among the types I-A, I-B and I-C in the case in which 
the bulk geometry does not belong to the type II, III or IV. For 
this purpose, we use Table 
\ref{table:SymmetryList}, which lists all higher symmetries that an 
$\IO(1)\times G(D-2,K)$-invariant spacetime can have. This list 
wasobtained by classifying all possible solutions to the 
Killingequation, but we do not give its derivation here, because it 
is ratherlengthy and tedious. In this table, the terms ``type II'', 
``typeIII'' and ``type IV'' have the meanings defined in the 
beginning ofthis section, while the term ``type I'' refers to all 
geometries other than those of these three types.

First, note that $G(D-2,\pm1)$ are semi-simple and do not have a 
non-trivial normal subgroup. In contrast, $G(D-2,0)=\IO(D-2)$ has 
the normal subgroup $\RF^{D-2}$, but the latter cannot be a normal 
subgroup of $\IO(1,1)$ or $\OG(2,1)=\OG(1,2)$ for $D\ge4$. Hence, 
from Table \ref{table:SymmetryList} we see that a subgroup of the 
type $G(D-2,K)$ contained in the isometry group of a spacetime that 
does not belong to the type II, III or IV is unique. This implies 
that brane configurations of the type I-A are mutually isometric and 
do not have a representation of the type I-B or I-C if the bulk 
geometry does not belong to the type II, III or IV. Further, it is 
also easy to see that configurations of the types I-B and I-C 
become mutually isometric only in the case in which the bulk 
spacetime has the structure $M_\lambda^{1,1}\times M_K^{D-2}$. 

Next, we comment on consequences for the other types obtained from 
Table \ref{table:SymmetryList}. First, the isometry group of a type 
III geometry is always $\IO(D-2,1)$. Hence, a configuration of the 
type I-C, if it exists, is not isometric to configurations of the 
type I-A. For the type IV, all configurations of a brane with the 
same $\sigma^2$ are mutually isometric. Finally, for a type II 
geometry, configurations of the types I-A and I-C are connected by 
isometries only when $U$ is constant, for which all configurations 
of the type I are mutually isometric.

\section{Summary and discussion}\label{sec:discussion}

In this paper, we have classified completely all  possible 
configurations of a brane, that is, a time-like hypersurface 
satisfying the condition $K_{\mu\nu}=\sigma g_{\mu\nu}$ in static 
$D$-dimensional spacetimes ($D\ge4$) with spatial symmetry 
$G(D-2,K)=\IO(D-2)(K=0)$, $\OG(D-1)(K=1)$ or $\OG_+(D-2,1)(K=-1)$, 
which has the bundle structure $(B^2,M_K^{D-2})$ with a 
2-dimensional base spacetime $B^2$ and a fibre $M_K^{D-2}$ with a 
constant curvature $K$. We summarize the main results in the 
following two theorems.

\begin{theorem} 
Configurations of a brane with $\sigma=0$ and allowed geometries are  
classified into the following three types: 
\begin{itemize} 
\item[I-A)] Brane configurations that are represented by subbundles   
$\Sigma=(B^2,F)$ of $M^D=(B^2,M_K^{D-2})$, where $F$ is a totally  
geodesic hypersurface $M$. Configurations of this type exist  
for any choice of $U,W$ and $S$, and are mutually  
isometric. Each configuration is invariant under $\IO(1)\times  
G(D-3,K')$ for some $K'\ge K$.  
\item[I-B)] $G(D-2,K)$-invariant configurations which are represented  
as $R=R(T)$ by solutions to 
\Eq{
R_T^2=WU(1-AU)\not=0;\ A>0.
}
In this case, the bulk geometry is restricted to a simple product $B^2\times  
M_K^{D-2}$, for which  $S=$const. 
\item[I-C)] Static configurations expressed as $R=$const in terms  
of solutions to 
\Eq{
U'=0,\quad S'=0.
}
Each configuration of this type corresponds to an $\IO(1)\times  
G(D-2,K)$-invariant totally geodesic hypersurface.  
\end{itemize} 
A brane with $\sigma=0$ can only have configurations of  
the type I-A in $\AdS^D$, while it can also have configurations of the 
types  I-B and I-C in $E^{D-1,1}$ and of the type I-B in 
$\dS^D$.  The latter are all mutually isometric in a given bulk 
geometry. In  a bulk spacetime that does not have constant 
curvature,  
configurations of the type I-B or I-C become isometric to those of  
the type I-A only when the bulk spacetime has the product structure  
$E^{0,1}\times M_K^{D-1}$.  
\end{theorem} 

Since the condition $\sigma=0$ is equivalent to the condition that 
the hypersurface is totally geodesic, this theorem gives the 
complete classification of totally geodesic time-like hypersurfaces 
in spacetimes with the $\IO(1)\times G(D-2,K)$ symmetry. 
From this point of view, the universal existence of configurations 
of the type I-A is rather trivial, because every totally geodesic 
hypersurface of a constant curvature space $R=$const is a 
fixed-point set for some involutive isometry that is also an 
isometry of the whole bulk spacetime. However, the $G(D-2,K)$ 
invariance of all the other configurations is a non-trivial result.

\begin{theorem} 
A brane with $\sigma\not=0$ can exist only for special bulk 
geometries.  These bulk geometries and corresponding brane 
configurations are classified into the following three types: 
\begin{itemize} 
\item[I-B)] Brane configurations that are $G(D-2,K)$-invariant and  
represented as $R=R(T)$ by solutions to  
\Eq{
R_T^2=U^2\left(1-\frac{U}{\sigma^2 R^2}\right)\not\equiv0.
}
In this case, the bulk geometry is restricted to those with $U=W$ and
$S=R$. For $S=R$, the condition $U=W$ is equivalent to the condition
that the Ricci tensor is isotropic in planes orthogonal to
$G(D-2,K)$-orbits. 
\item[I-C)] Static and $G(D-2,K)$-invariant brane
configurations expressed as $R=$const in terms of solutions to
\Eq{
\frac{U'}{U}=\frac{2S'}{S},\quad
W\left(\frac{S'}{S}\right)^2=\sigma^2.
}
\item[II)] Static brane configurations in the bulk geometries with 
metrics  of the form 
\Eq{ 
ds_D^2=d\sigma_{\lambda,D-1}^2 - UdT^2,
}
where $d\sigma_{\lambda,D-1}^2$ is the metric of a $(D-1)$-dimensional
constant curvature space $M_\lambda^{D-1}$ with sectional curvature
$\lambda$. In this case, $U$ is a function on this space that is
invariant under a subgroup $G(D-2,K)$ of the isometry group of
$M_\lambda^{D-1}$. Allowed forms of $U$ and brane configurations are
given as follows:
\begin{itemize} 
\item[i)] $\lambda=0$: In a Cartesian coordinate system $\bm{x}$ for  
$E^{D-1}$, $U=((\bm{x}-\bm{a})^2+k)^2$ and brane configurations are   
represented as $(\bm{x}-\bm{b})^2=1/\sigma^2$ with  
$(\bm{b}-\bm{a})^2=1/\sigma^2-k$. 
\item[ii)] $\lambda=1/\ell^2>0$: In a homogeneous coordinate system  
$X$ in which $S^{D-1}$ is expressed as $X\cdot X=1$, $U=(P\cdot  
X+k)^2$ ($k\not=0$) and brane configurations are represented as  
$Q\cdot X=1$ with $Q\cdot P=-k$ and $Q\cdot Q=1+1/(\ell\sigma)^2$. 
\item[iii)] $\lambda=-1/\ell^2<0$: In a homogeneous coordinate  
system $Y$ in which $H^{D-1}$ is expressed as $Y\cdot Y=-1$,  
$U=(P\cdot Y +k)^2$ ($k\not=0$) and brane configurations are  
represented as $Q\cdot Y=1$ with $Q\cdot P=k$ and $Q\cdot  
Q=1/(\ell\sigma)^2-1$.  
\end{itemize} 
Although these brane configurations are not $G(D-2,K)$-invariant, they 
are still $G(D-3,K')$-invariant for some $K'\ge K$ and mutually isometric. 
\end{itemize}
Configurations of the type I-B and those of the type I-C are 
isometric only when the bulk spacetime is a product of a 
2-dimensional constant curvature spacetime and a constant curvature 
space $M_K^{D-2}$, provided that it is not a constant curvature 
spacetime. In constant curvature spacetimes, a brane with 
$\sigma\not=0$ can have only configurations of the type I-B, except 
for a brane with $\ell^2\sigma^2=1$ in $\AdS^D(\ell)$, for which 
only 1-C configurations are allowed.
\end{theorem}

Here, note that type I-B configurations can have 
different isometry classes only when a type I-C configuration 
exists. Since a type I-C configuration does not exist in a generic 
spacetime, this implies that configurations of a brane with 
$\sigma\not=0$ are mutually isometric in a generic type I-B geometry.

Now we give some comments on implications of these results. First, 
as stated in the Introduction, one purpose of the present paper has been 
to extend the analysis done by Chamblin, Hawking and Reall on the 
possibility of a black hole geometry being induced on a vacuum brane from a 
(pseudo-)spherically symmetric bulk spacetime. From this point of 
view, only configurations of a brane with spatial symmetry lower 
than that of the bulk are relevant. However, our results show that 
such configurations are static and possible only in bulk spacetimes 
of the type II with the special forms of $U$ given above. It is clear 
that these solutions do not give a black hole geometry on the brane, 
because the space has a constant curvature for them. Further, 
although $U$ can vanish when $k$ has the appropriate sign, the 
spacetime curvature diverges at points where $U$ vanishes. Hence, we 
can conclude that we can never obtain a brane configuration 
representing a vacuum black hole from a static spacetime solution to 
the Einstein equations with the assumed spatial symmetry, 
irrespective of the bulk matter content.

Next, we comment on mathematical consequences. First, our results
extend the theorem concerning everywhere umbilical hypersurfaces in
Euclidean spaces mentioned in the Introduction to the case of constant
curvature spacetimes. Our results also extend this theorem to
non-constant curvature spacetimes, in the sense that spatial sections
of a brane with $\sigma\not=0$ always respect the spatial symmetry
$G(D-2,K)$ of a spacetime, and a brane configuration is always
$G(D-3,K')$-invariant with $K'\ge K$. Our results also provide
extensions of the famous rigidity theorem on hypersurfaces in
Euclidean spaces, which states that two embeddings of a Riemannian
manifold into a Euclidean space as hypersurfaces become isometric if
the extrinsic curvatures coincide.\cite{Kobayashi.S&Nomizu1963B} For
example, our results on a brane in constant curvature spacetimes give
a direct extension of this theorem to time-like hypersurfaces with
$K_{\mu\nu}=\sigma g_{\mu\nu}$ in constant curvature
spacetimes. Further, our results imply that the non-existence of a
type I-C solution is a necessary condition for the same rigidity to
hold for hypersurfaces with $\sigma\not=0$ in spacetimes that do not
have a constant curvature.

\section*{Acknowledgment}

The author thanks the participants of the 2nd workshop ``Braneworld
 -- Dynamics of spacetime with boundary'' held at YITP in January, 
2002 for useful comments. This work is supported by the JSPS Grant 
No.11640273.

\appendix

\section{Geometric Quantities for the Bulk 
Geometry}\label{appendix:bulkgeometry:formulas}

In this appendix we list formulas for the Christoffel symbols and the 
curvature tensor of the bulk metric \eqref{bulkgeometry:symmetric}.


For the bulk metric \eqref{bulkgeometry:symmetric}, the non-vanishing 
components of the Christoffel symbols are given by
\Eqrsubl{Christoffel:bulk}{
&& \tilde \Gamma^T_{RT}=\frac{U'}{2U},\\
&& \tilde \Gamma^R_{TT}=\frac{1}{2}WU',\ 
\tilde \Gamma^R_{RR}=-\frac{W'}{2W},\ 
\tilde \Gamma^R_{ij}=-SS'W\gamma_{ij},\\
&& \tilde \Gamma^i_{jR}=\frac{S'}{S}\delta^i_j,\ 
\tilde \Gamma^i_{jk}=\Gamma^i_{jk},
}
where the prime denotes differentiation with respect to the argument
of the corresponding function, and $\Gamma^i_{jk}$ is the Christoffel
symbol for $\gamma_{ij}$, which is expressed in terms of $\rho(\chi)$
and the Christoffel symbol $\hat\Gamma^A_{BC}$ for the metric
$\hat\gamma_{AB}$ as
\Eq{
\Gamma^\chi_{AB}=-\rho\rho'\hat\gamma_{AB},\ 
\Gamma^A_{B\chi}=\frac{\rho'}{\rho}\delta^A_B,\ 
\Gamma^A_{BC}=\hat\Gamma^A_{BC}.
\label{Christofel:H_K}
}
%


The curvature tensors have simpler expressions when we write the bulk 
metric as
\Eq{
ds_{D}^2=g_{ab}dx^adx^b + S(R)^2\gamma_{ij}dz^i dz^j.
}
With this notation, the Riemann curvature tensor is given by
\Eqrsubl{Riemann:bulk}{
&& \tilde R_{abcd}=\frac{1}{2}{}^2\!R (g_{ac}g_{bd}-g_{ad}g_{bc}),\\
&& \tilde R^i{}_{ajb}=\left(\frac{S'}{S}\tilde\Gamma^R_{ab}
-\frac{S''}{S}\delta^R_a\delta^R_b\right)\delta^i_j,\\
&& \tilde 
R_{ijkl}=(K-S'{}^2W)S^2
(\gamma_{ik}\gamma_{jl}-\gamma_{il}\gamma_{jk}),}
and the Ricci tensor is given by
\Eqrsubl{Ricci:bulk}{
&& \tilde R_{ab}=\frac{1}{2}{}^2\!R g_{ab} 
+ (D-2)\left(\frac{S'}{S}\tilde\Gamma^R_{ab}
-\frac{S''}{S}\delta^R_a\delta^R_b\right),\\
&& \tilde R_{ai}=0,\\
&& \tilde R^i_j=\left[-\frac{(UWS'{}^2)'}{2USS'}
+(D-3)\frac{K-S'{}^2W}{S^2}\right]\delta^i_j,\\
&& \tilde R={}^2\!R
 -(D-2)\frac{(UWS'{}^2)'}{USS'}+(D-2)(D-3)\frac{K-W}{S^2},
}
where
\Eq{
{}^2\!R=-\frac{1}{2U'}\left(\frac{WU'{}^2}{U}\right)'
=-\frac{W}{U}\left(U''+\frac{W'U'}{2W}
-\frac{(U')^2}{2U}\right).
}
%

\section{$\partial_T\partial_iu=f(T,\bm{z})\partial_Tu \partial_i u$}
\label{appendix:DTDiu}

In this appendix, we show that if $u(T,\bm{z})$ 
($\bm{z}=(z^1,\cdots,z^n)$) satisfies the equation
\Eq{
\partial_T\partial_iu=f(T,\bm{z})\partial_T u \partial_i u,
}
$u$ can be written as $u=F(T,X(\bm{z}))$ in terms of some function 
$X(\bm{z})$.

First, note that from this equation it immediately follows that for 
a fixed value of $T$, $\partial_T u$ is constant on a hypersurface 
in the $z$ space on which $u$ is constant. Hence, $u_T$ can be written 
in terms of some function $G(T,u)$ as $\partial_Tu=G(T,u)$. Let the 
solution of the ordinary differential equation $dv/dT=G(T,v)$ for 
the initial condition $v(T_0)=v_0$ be $v=F(T,v_0)$. Then, for any 
function $X(\bm{z})$, we obtain
\Eq{
\partial_T F(T,X(\bm{z}))=G(T,F(T,X(\bm{z}))),
}
and $F(T_0,X(\bm{z}))=X(\bm{z})$. Since the solution of the equation 
$\partial_T u=G(T,u)$ is uniquely determined if 
$u(T_0,\bm{z})$ is given, this implies that any solution $u$ to the 
equation $\partial_T u=G(T,u)$ can be written $u=F(T,X(\bm{z}))$ with 
$X(\bm{z})=u(T_0,\bm{z})$.

\section{$D_iD_ju=\alpha g_{ij} +\beta D_iuD_ju$}\label{appendix:DDu}

In this appendix, we give the general solution to the equation
\Eq{
D_iD_ju=\alpha g_{ij}+\beta D_iuD_ju
\label{ConformalEq:general}}
on an $n$-dimensional constant curvature space $M_K^n$ with sectional 
curvature $K$. Here, $D_i$ is the covariant derivative with respect to 
the metric $g_{ij}$ of $M_K^n$, and $\alpha$ and $\beta$ are functions 
of $u$. We assume that $n>1$.

First, we show that the problem can be reduced to the case $\beta=0$. 
Let us change the unknown function $u$ to $v$ by $u=f(v)$. Then, 
\eqref{ConformalEq:general} is transformed into
\Eq{
f' D_iD_jv =\alpha g_{ij} + \left((f')^2\beta-f''\right)D_ivD_jv.
}
Hence, if $f$ is so chosen as to satisfy the equation $f''=\beta(f) 
(f')^2$, i.e., 
\Eq{
v=f^{-1}(u)=\int du \exp -\int^u \beta(u')du',
}
we obtain an equation of the type \eqref{ConformalEq:general} with 
$\beta=0$. 

For $\beta=0$, the consistency of \eqref{ConformalEq:general} gives
\Eq{
D_i\alpha=D^jD_iD_j u=R^j_iD_ju + D_i\triangle u
=R^j_iD_j u + n D_i\alpha.
}
Here, since $M_K^n$ is a constant curvature space, its Ricci tensor 
$R_{ij}$ is  
\Eq{
R_{ij}=(n-1)Kg_{ij}.
}
Hence, we obtain
\Eq{
D_i(\alpha+Ku)=0,
}
whose general solution is 
\Eq{
\alpha(u)=-Ku+\alpha_0.
}
If $\alpha$ does not have this form, \eqref{ConformalEq:general} with 
$\beta=0$ has no solution.

First, we consider the  case  $K=0$, in which we can choose a 
Cartesian coordinate system $z^i$ for which 
$g_{ij}=\delta_{ij}$. In this coordinate system, the equation can be 
easily integrated. The general solution is given by
\Eq{
u=\frac{1}{2}\alpha_0 z\cdot z + a_i z^i + b,
\label{u:K=0}}
where $a^i$ and $b$ are constants.

Next, in the case $K=\pm1$, by the replacement $u-K\alpha_0 \tend u$, 
the problem is reduced to that of the case $\alpha=-Ku$,
\Eq{
D_iD_j u=-Kg_{ij} u.
\label{ConformalEq}
}
Then, it follows from this equation that 
\Eq{
D_i[(Du)^2+Ku^2]=2(D_iD_ju + Ku g_{ij})D^ju=0,
}
where $(Du)^2=g^{ij}D_iuD_ju$. Hence, 
\Eq{
(Du)^2+Ku^2=c \ (\text{constant}).
\label{(Du)^2}}
Further, for any curve $z^i=z^i(t)$, \eqref{ConformalEq} gives the 
ordinary differential equation
\Eq{
\dot u= \dot z^i u_i,\ 
\dot u_i=\Gamma^j_{ik}\dot z^k u_j-Ku g_{ij}\dot z^j.
}
This implies that $u$ is uniquely determined if the values of $u$ 
and $u_i=D_i u$ at some point are specified. In particular, $(Du)^2$ 
can vanish only at isolated points, and each level set of $u$ is a 
smooth hypersurface, except at such points, for any non-trivial 
solution $u$. 

Now, let us calculate the extrinsic curvature $K_{AB}$ of the $u=$ 
const surface $\Sigma_u$. Since the unit normal to $\Sigma_u$ is 
given by $n_A=D_A u /|Du|$, from \eqref{ConformalEq}, $K_{AB}$ can be  
expressed as
\Eq{
K_{AB}=-D_A n_B= \frac{Ku}{|Du|}g_{AB},
}
which implies that $\Sigma_u$ gives an everywhere umbilical 
hypersurface. Hence, from the argument in \S\ref{sec:EUHinCCS}, $u$ 
can be expressed as a function of $v$ for an appropriate choice of 
$v$ giving the decomposition \eqref{CCS:metric:1}. 

Now, using this observation, we derive expressions for $u$ in terms 
of homogeneous coordinate systems for $M_K^n$. First, for $K=1$, the 
metric of the unit sphere $S^n$ embedded in $E^{n+1}$ can be expressed 
in terms of a homogeneous coordinate system $\Omega^a$ as
\Eq{
ds^2=d\Omega\cdot d\Omega;\ 
\Omega\cdot\Omega=1.
}
From the observation above and the argument in 
\S\ref{sec:TypeII:S}, $u$ must be a function of $X=P \cdot \Omega$ 
for some fixed constant vector $P=(P^a)$. Hence, \eqref{ConformalEq} 
is written
\Eq{
u_X D_iD_jX + u_{XX}D_iXD_jX=-g_{ij}u.
}
However, it is easily checked that each $\Omega^a$ satisfies 
\eqref{ConformalEq} with $K=1$. This implies that $u(X)$ satisfies 
$Xu_X=u$ and $u_{XX}=0$. Therefore, the general solution to 
\eqref{ConformalEq} with $K=1$ is  
\Eq{
u=P\cdot \Omega;\ c=P\cdot P,
\label{u:K=1}}
where we have used the relation \eqref{Id:S} to calculate $c$. Through
an appropriate transformation in $\OG(n+1)$, $u$ can be written
$u=\sqrt{c}\Omega^{n+1}$. If we introduce the angular coordinate
$\chi$ by $\Omega^{n+1}=\cos\chi$, $u$ is given by 
\Eq{
u=\sqrt{c}\cos\chi.
}

The derivation of the corresponding formula for $K=-1$ is almost the 
same. The metric of the unit hyperboloid $H^n$ embedded in $E^{n,1}$ 
is expressed as
\Eq{
ds^2=dY\cdot dY;\ 
Y\cdot Y=-1,
}
where each homogeneous coordinate $Y^\mu$  satisfies 
\eqref{ConformalEq}. Hence, by the same reasoning as that applied in
the case $K=1$, we can conclude that the general solution to
\eqref{ConformalEq} with $K=-1$ is 
\Eq{
u=P\cdot Y;\ c=P\cdot P,
\label{u:K=-1}}
where we have used the relation \eqref{Id:H} to calculate $c$.

We can also express this solution in terms of the coordinate system 
$(\chi,\theta,\cdots)$ used in \eqref{metric:M_K}, which is related 
to the homogeneous coordinate system $Y$ by
\Eq{
Y^0=\cosh\chi,\ 
Y^i=\sinh\chi \Omega^i.\ (i=1,\cdots,n)
}
The result depends on the sign of $c$. First, for $c<0$, we can put 
\eqref{u:K=-1} into the form $u=\sqrt{|c|}Y^0$ by a transformation in 
$\OG_+(n,1)$. Then, $u$ is given by 
\Eq{
u=\sqrt{|c|}\cosh\chi.
}
Second, for $c=0$, we can put \eqref{u:K=-1} into the form 
$u=k(Y^0-Y^n)$ by a Lorentz transformation. Hence, $u$ can be 
expressed in terms of 
$\chi$ and $\theta$ as
\Eq{
u=k(\cosh\chi-\sinh\chi\cos\theta).
}
Finally, for $c>0$, we can put \eqref{u:K=-1} into the form  $u=\sqrt{c}Y^n$, 
which reads
\Eq{
u=\sqrt{c}\sinh\chi\cos\theta.
}
%


\end{document}